\documentclass{ws-procs961x669}            
\begin{document}
\title{The self-gravitating Fermi gas in Newtonian gravity\\
 and general
relativity}

\author{Pierre-Henri Chavanis$^*$}

\address{Laboratoire de Physique
Th\'eorique, Universit\'e de Toulouse,
CNRS, UPS, France\\
$^*$E-mail: chavanis@irsamc.ups-tlse.fr}

\begin{abstract}
We review the history of the self-gravitating Fermi gas in Newtonian gravity and
general relativity. We mention applications to white dwarfs, neutron stars
and dark matter halos. We describe the nature of instabilities and phase
transitions in the self-gravitating Fermi gas as energy (microcanonical
ensemble) or temperature (canonical ensemble) is reduced. When
$N<N_{\rm OV}$, where $N_{\rm OV}$ is the Oppenheimer-Volkoff critical particle
number, the self-gravitating Fermi gas experiences a gravothermal catastrophe
at $E_c$ stopped by quantum mechanics (Pauli's exclusion principle). The
equilibrium state has a core-halo structure made of a quantum core
(degenerate fermion ball) surrounded by a classical isothermal halo. When
$N>N_{\rm OV}$, a new turning point appears at an energy $E''_c$ below
which the system experiences a gravitational collapse towards a black hole
[P.H. Chavanis, G. Alberti, Phys. Lett. B {\bf 801}, 135155 (2020)].
When $N_{\rm OV}<N<N'_*$, the self-gravitating Fermi gas experiences a
gravothermal catastrophe at $E_c$ leading to a fermion ball, then a
gravitational collapse at $E''_c$ leading  to a black hole. When $N>N'_*$,
the condensed branch disappears and the instability at  $E_c$ 
directly leads to a black hole. We discuss implications of these
results for
dark matter halos made of massive neutrinos.
 
\end{abstract}


\bodymatter

\section{Introduction}

The self-gravitating Fermi gas can have applications in different astrophysical
systems ranging from white dwarfs and neutron stars to dark matter halos,
where
the fermions are electrons, neutrons and massive neutrinos respectively. The
study of the self-gravitating Fermi gas is also of fundamental conceptual
importance as it combines quantum mechanics and general relativity. Initially,
fermionic models were developed at zero temperature ($T=0$) but they have been
later
generalized at nonzero temperature, especially in the case of dark matter
halos. In these Proceedings, we provide a brief history of the
self-gravitating Fermi gas. A more detailed historical account of the
statistical mechanics and
thermodynamics of self-gravitating systems (classical and quantum) in Newtonian
gravity and general relativity can be found in Refs. \cite{acb,acf,gr1,gr2}.

The statistical equilibrium state of a system of self-gravitating
fermions can be determined from a maximum entropy principle.  For systems with
long-range interactions the mean field
approximation becomes exact in an appropriate thermodynamic
limit.\cite{campabook,gr1} The most probable
distribution of an isolated system of self-gravitating fermions at statistical
equilibrium is obtained by
maximizing the Fermi-Dirac entropy $S$ at fixed mass-energy ${\cal E}=Mc^2$ and
particle number $N$:
\begin{equation}
\label{i1}
 \max \left\lbrace S \quad | \quad {\cal E}=Mc^2, \quad N \quad
{\rm
fixed}
\right\rbrace.
\end{equation}
The variational problem for the first variations reads
\begin{equation}
\label{i2}
{\delta S}/{k_B}-\beta_{\infty}\delta {\cal E}+\alpha\delta N=0,
\end{equation}
where $\beta_{\infty}=1/k_B T_{\infty}$ and $\alpha=\mu_{\infty}/k_B T_{\infty}$
are Lagrange multipliers
associated with the conservation of mass-energy and particle number. Here,
$T_{\infty}$ and $\mu_{\infty}$ represent the temperature and the chemical
potential measured by an observer at
infinity. The maximization problem (\ref{i1}) is associated with the
microcanonical ensemble. If the system is in
contact with a thermal bath fixing the temperature  $T_{\infty}$ the statistical
equilibrium state is obtained by minimizing the free energy $F={\cal
E}-T_{\infty} S$ at fixed particle number $N$:
\begin{equation}
\label{i3}
\min \left\lbrace F={\cal E}-T_{\infty} S \quad | \quad N\quad
{\rm
fixed}
\right\rbrace.
\end{equation}
This minimization problem is associated with the canonical ensemble. At $T=0$
the equilibrium state is obtained by minimizing the mass-energy ${\cal E}=Mc^2$
at fixed particle number $N$. The equilibrium states in the microcanonical and
canonical ensembles are the same. They are determined by the variational
principle (\ref{i2}). However, their stability may differ in the two ensembles.
This is the notion of ensemble inequivalence for systems with long-range
interactions.\cite{gr1,campabook} Microcanonical stability implies canonical
stability but the
converse is
wrong.

The equilibrium state of a gas of self-gravitating fermions results from the
balance between the repulsion due to the quantum pressure (Pauli's exclusion
principle) and the gravitational attraction. The variational principle
(\ref{i2}) yields all the equations that
we need to determine the equilibrium state of the self-gravitating Fermi gas:
(i) the Fermi-Dirac distribution function; (ii) the ideal equation of
state of fermions; (iii) the Oppenheimer-Volkoff
equations determining the condition of hydrostatic equilibrium in
general relativity; (iv) the Tolman-Klein relations 
expressing how the local temperature $T(r)$ and the local chemical potential
$\mu(r)$ are affected 
by
the metric. We can solve these equations numerically and  plot the caloric
curve $T_{\infty}({\cal E})$ relating the temperature to the energy. When
$T>0$ we need to enclose the system within a spherical box of
radius $R$ in order to prevent its evaporation and have equilibrium
states with a finite mass. In the general case, the caloric curve depends on
$N$ and $R$. For  
convenience, instead of $T_{\infty}({\cal E})$, we shall plot
$\beta_{\infty}(-E)$ where $E=(M-Nm)c^2$ is the
binding energy which reduces to the usual energy $E=K+W$ (kinetic $+$ potential)
in the
nonrelativistic limit $c\rightarrow +\infty$. At $T=0$, the system is
self-confined (without the need of a box) and we shall plot the mass-radius
relation $M(R)$ where $R$ denotes here the radius where the density vanishes.
The maximum
entropy formalism for classical and quantum self-gravitating systems in
Newtonian gravity and general relativity is reviewed in Refs. \cite{gr1,gr2}
where all the equations are derived and an exhaustive list of references is
given.

\section{Self-gravitating fermions at $T=0$}
\label{sec_zero}

The study of a self-gravitating gas of fermions started in the
context of white dwarf stars when Fowler \cite{fowler} first realized that these
compact objects owe their stability to the quantum pressure of the degenerate
electron gas. Indeed, the
quantum pressure arising from the Pauli exclusion principle is able to
counteract the gravitational attraction and explain the very high densities of
white dwarf stars.
Early studies were devoted to determining the ground state ($T=0$) of the
system. Nonrelativistic white dwarf stars are equivalent to a polytropic gas of
index $n=3/2$. Their density profile
can be obtained by solving the  Lane-Emden equation numerically.
The density profile of white dwarf stars at $T=0$ 
has a compact support, i.e., the density vanishes at a finite radius. The
mass-radius relation of nonrelativistic white dwarf stars was first obtained by
Stoner
\cite{stoner29}, Milne \cite{milne} and Chandrasekhar
\cite{chandra31nr}.\footnote{Stoner
\cite{stoner29} developed an
analytical approach based on a uniform density approximation for the star
while Milne \cite{milne} and Chandrasekhar
\cite{chandra31nr} developed a numerical approach based on the Lane-Emden theory
of
polytropes.\cite{emden}}
 They
showed that the
radius of the star decreases as the mass increases according to the law
$M=91.9\, \hbar^6/(G^3m^8R^3)$ (see Fig. \ref{ftz}-a).\cite{chandrabook} All
the
configurations
of the series of equilibria are
stable.

\def\figsubcap#1{\par\noindent\centering\footnotesize(#1)}
\begin{figure}[h]%
\begin{center}
\parbox{2.1in}{\includegraphics[width=2in]{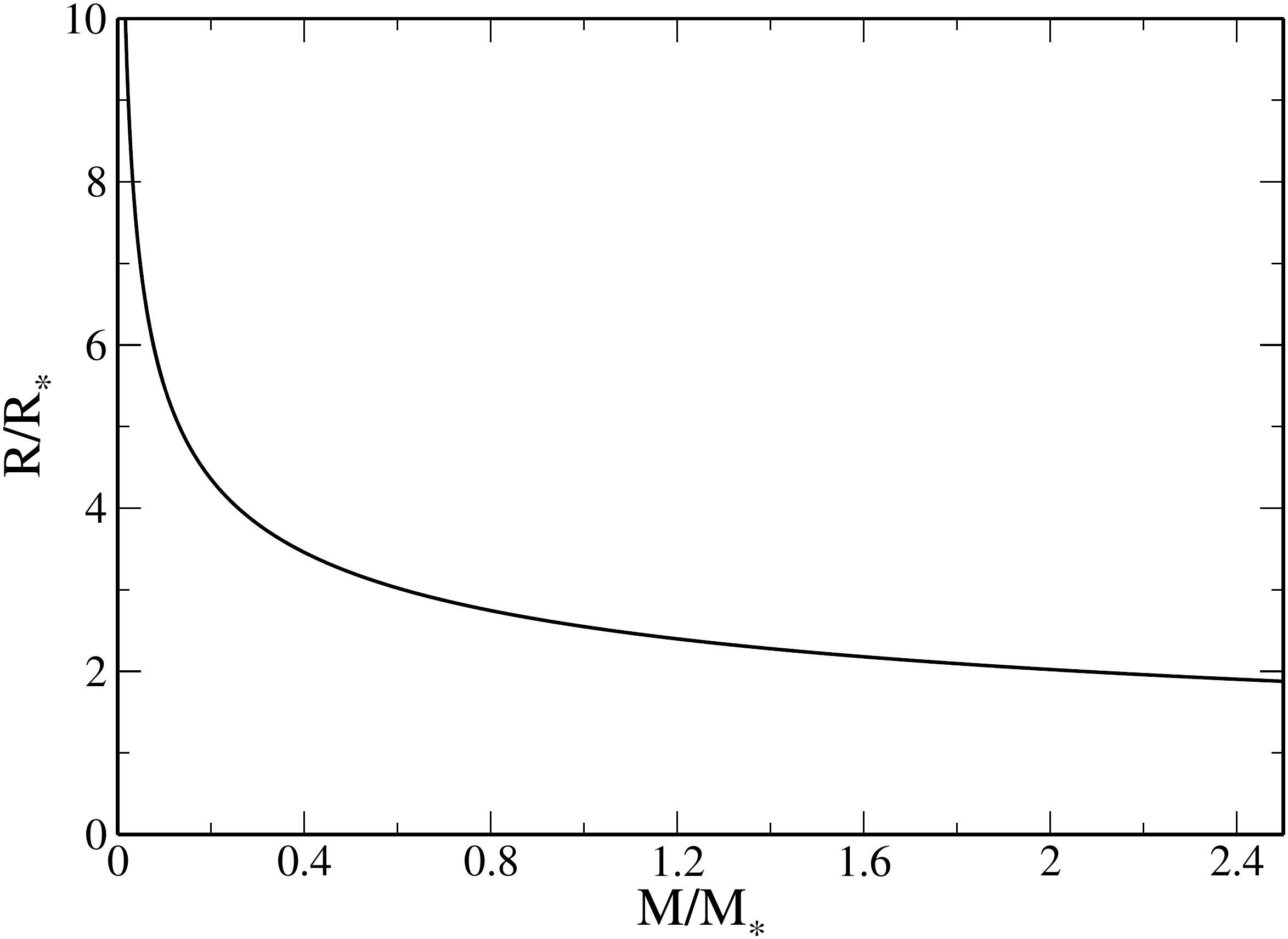}\figsubcap
{a}}
  \hspace*{4pt}
  \parbox{2.1in}{\includegraphics[width=2in]{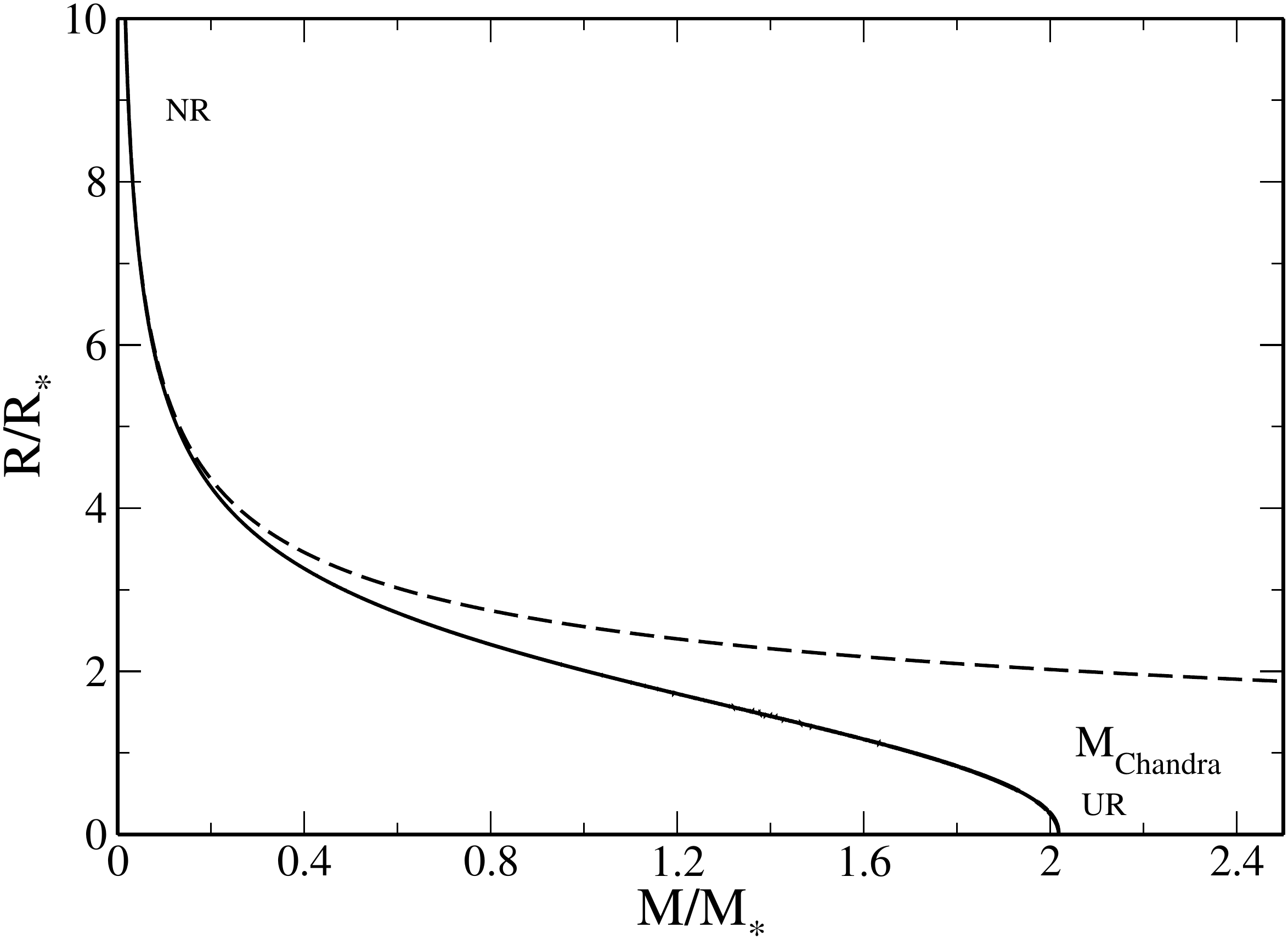}\figsubcap{b}}
  \caption{Mass-radius relation of self-gravitating fermions at $T=0$ in
Newtonian gravity (the mass
is normalized by $M_*=(3\pi/4)^{1/2}\hbar^{3/2}c^{3/2}/(G^{3/2}m^2)$ and the
radius by
$R_*=(3\pi/4)^{1/2}\hbar^{3/2}/(c^{1/2}G^{1/2}m^2)$). (a) Nonrelativistic white
dwarf stars.
(b) Special relativistic white dwarf stars.}%
  \label{ftz}
\end{center}
\end{figure}

The fact that relativistic effects become important in white dwarf stars whose
mass is of the order of the solar mass was first reported by Frenkel
\cite{frenkel} in a not well-known paper. However, he did not consider the
implications of this result. Special relativistic effects in white dwarf stars
were
studied in detail by Anderson,\cite{anderson} Stoner,\cite{stoner30}
Chandrasekhar,\cite{chandra31} and Landau.\cite{landau32} They found that no
equilibrium state is possible above a maximum mass, now known as the
Chandrasekhar limit.\footnote{See the introduction of Ref.\cite{wddimd} for a
short
history of the discovery of the maximum mass of white dwarf stars.} These
authors considered the equation of state of a relativistic Fermi gas at $T=0$
and used Newtonian gravity appropriate to white dwarf
stars.\footnote{In principle, general relativistic effects
become important
close to the Chandrasekhar maximum mass\cite{kaplan1949,chtoop64}
but other phenomena like Coulomb corrections to the electron pressure and the
formation of neutrons by inverse beta decay destabilize the star before
general relativistic effects come into play.\cite{hs,htww65}} An
ultrarelativistic
Fermi gas at $T=0$ is equivalent to
a polytrope of index $n=3$. Its density profile is obtained by solving the
corresponding Lane-Emden equation and it has a compact support. For a polytrope 
$n=3$, the mass-radius relation
degenerates and indicates that different configurations with an arbitrary radius
can exist at the same mass $M_{\rm Chandra}=3.1\, M_P^3/m^2=1.5\, M_{\odot}$,
where
$M_P=(\hbar c/G)^{1/2}$  is the Planck mass and $m$ is the proton mass. This
argument immediately implies the existence of a critical mass.\cite{chandra31}
In a more
detailed study, 
Chandrasekhar \cite{chandra35} considered partially relativistic
configurations and numerically obtained the complete mass-radius relation of
white dwarf stars, valid for arbitrary densities,  joining the nonrelativistic
limit to the ultrarelativistic one (see Fig. \ref{ftz}-b).\footnote{A similar
mass-radius relation was
obtained earlier by Stoner \cite{stoner30} from an approximate analytical
model based on uniform density stars.} As $M$ approaches $M_{\rm Chandra}$ the
radius of the star
tends to zero while its density tends to infinity, leading to a Dirac peak.
This
study unambiguously shows the absence of equilibrium state above a maximum mass.
Therefore, the quantum pressure arising from the Pauli exclusion principle
cannot balance the gravitational attraction anymore when the star becomes
sufficiently relativistic (or when its mass is too large). This is a striking
effect of relativity combined with quantum mechanics and gravity. However, the
result of Chandrasekhar
\cite{chandra35} was severely criticized by Eddington \cite{eddington35} who
argued that the absence of equilibrium states above a maximum mass leads to a
{\it reductio ad absurdum} of the formula of relativistic degeneracy. Although
the arguments of Eddington were entirely unfounded, his enormous prestige led to
an early rejection of Chandrasekhar's work by many in the astronomical
community. This pushed Chandrasekhar to abandon the subject, and delayed the
discovery of the phenomenon of gravitational collapse and the concept of black
hole.

\begin{figure}[h]
\begin{center}
\includegraphics[width=2.5in]{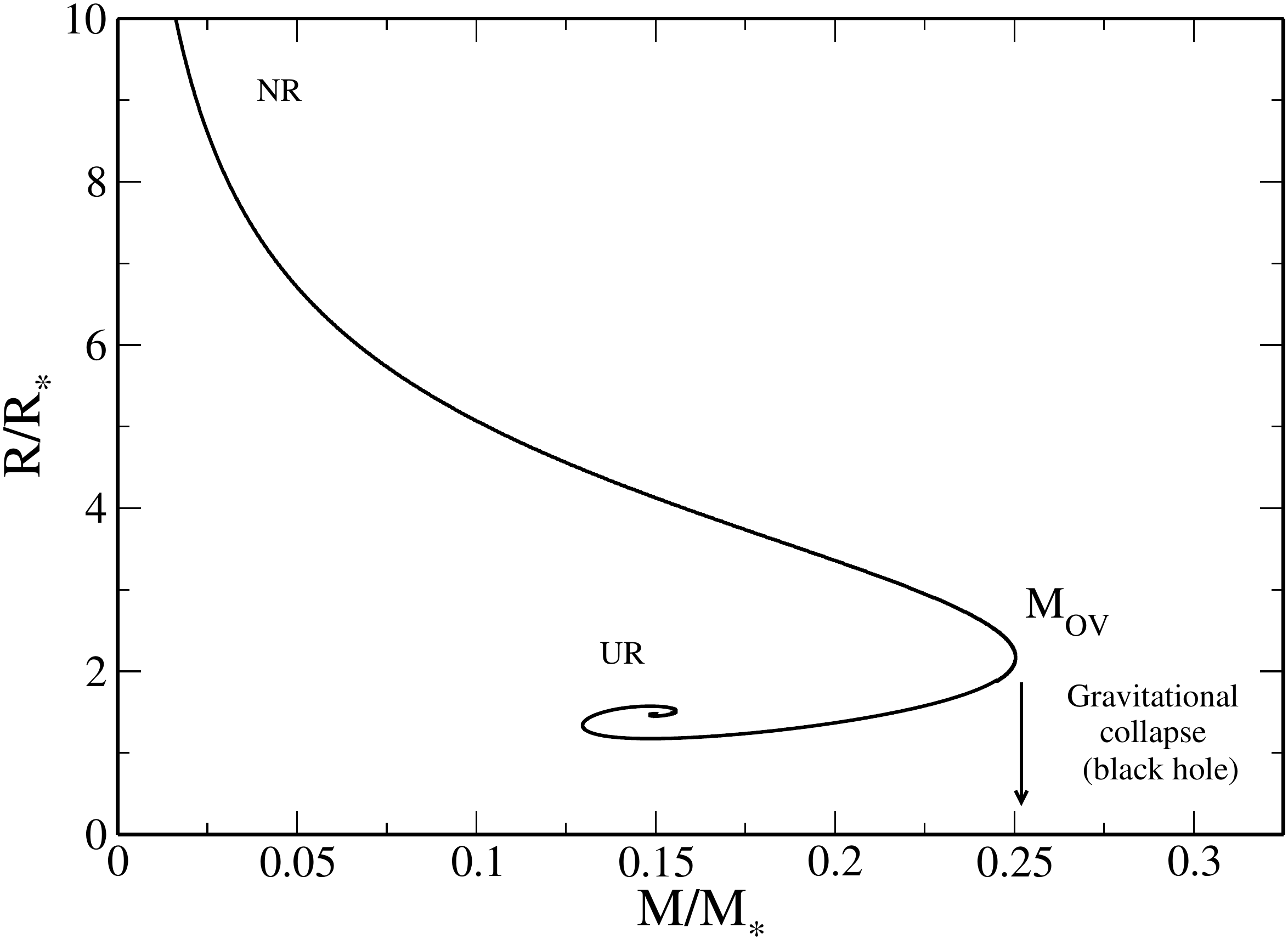}
\end{center}
\caption{Mass-radius relation of self-gravitating fermions at
$T=0$ in general relativity (neutron stars).}
\label{courbeOV}
\end{figure}

In the following years, similar results were found by
Oppenheimer and Volkoff
\cite{ov} in
connection to neutron stars. They solved the equations of general relativity
with the relativistic equation of state for fermions at $T=0$ and found that the
mass-radius relation of neutron stars presents a turning point of
mass (see Fig. \ref{courbeOV}).\footnote{It was shown
later that the mass-radius relation of neutron stars forms a spiral and that a
mode of stability is lost at each turning point of mass.\cite{htww65}} As a
result, no equilibrium state exists above a maximum mass $M_{\rm OV}=0.384\,
M_P^3/m^2=0.710\, M_{\odot}$, where $m$ is the neutron mass,  called the
Oppenheimer-Volkoff limit (note that the density profile with the maximum mass
$M_{\rm OV}$  is {\it not} singular contrary to the Newtonian density profile at
the
maximum mass $M_{\rm Chandra}$). They argued that, above that mass, the star
undergoes
gravitational collapse. This problem was specifically studied by Oppenheimer and
Snyder \cite{os} who obtained  an analytical solution of the Einstein equations
describing the collapse of a pressureless gas up to its Schwarzschild radius.
Strangely enough, these important results did not receive much attention until
the 1960's. At that epoch, detailed models of compact objects with more
realistic equations of state taking into account
the repulsive effect of nuclear forces and connecting white dwarfs to neutron
stars
were
constructed and the fundamental discoveries of Chandrasekhar, Landau and
Oppenheimer and Volkoff were confirmed (unfortunately, the early contributions
of Anderson and
Stoner were forgotten).\cite{htww65} Pulsars were discovered by
Hewish {\it et al.} \cite{hewish}
in 1968. The
same year, Gold \cite{gold1,gold2} proposed that pulsars are rotating
neutron stars, and this is generally accepted today. It is also at that moment
that the name
``black hole'' was used by Wheeler
\cite{wheelerBH} to designate the object resulting from gravitational
collapse, and became
popular.

\section{Nonrelativistic classical  particles at $T>0$ and self-gravitating
radiation}
\label{sec_nrc}

The thermodynamics of self-gravitating systems is a
fascinating subject.\cite{paddy,found,ijmpb} Its study started
with the pioneering work of Antonov
\cite{antonov} who considered
an isolated system of nonrelativistic classical particles in gravitational
interaction. He used a microcanonical ensemble
description in which the mass and the energy are conserved. This situation
applies approximately  to stellar systems like globular clusters. Their
equilibrium state (most probable state) can be obtained by maximizing the
Boltzmann
entropy $S$ at fixed mass $M$ and energy $E$.\cite{ogorodnikov}  This leads to
the mean
field Boltzmann distribution which is self-consistently coupled to the
Poisson equation.
The Boltzmann-Poisson equation was previously introduced
and studied in the
context of isothermal stars.\cite{emden,chandrabook} It can be
reduced to the Emden equation that has to be solved numerically. Antonov
\cite{antonov} observed
that  no maximum entropy state exists  in an infinite domain (the solution of
the Emden equation has an infinite mass), so he proposed to confine the
particles within a spherical box of radius $R$. This artifice prevents the
evaporation of the system and leads to a well-defined mathematical problem. By
computing the second variations of entropy, Antonov \cite{antonov} showed that
equilibrium
states with a density contrast ${\cal R}=\rho_0/\rho(R)<709$, where $\rho_0$ is
the central density and $\rho(R)$ the density on the edge of the box, are
thermodynamically stable (entropy maxima) while  equilibrium states with a
density contrast ${\cal R}>709$ are thermodynamically unstable (saddle points of
entropy). Lynden-Bell and Wood \cite{lbw} rediscussed the results of Antonov
\cite{antonov} in more physical terms. They plotted the series of equilibria
$E({\cal R})$ and showed that it displays damped oscillations. As a result,
there is no equilibrium state with an energy $E<E_c=-0.335\, GM^2/R$, where
$E_c$ corresponds to the first turning point of energy (with a density contrast
${\cal R}_c=709$). Invoking the Poincar\'e turning point
criterion,\cite{poincare}  they concluded that the series of equilibria becomes
unstable
at the minimum energy $E_c$. In this manner, they recovered the critical density
contrast ${\cal R}_c=709$ found by Antonov.\cite{antonov} They also
interpreted
the Antonov instability in terms of a ``gravothermal catastrophe'' caused by the
negative specific heat of the system in its densest parts. When this
instability occurs, the system undergoes core collapse. This ultimately leads to
a binary
star surrounded by a hot halo. Lynden-Bell and Wood
\cite{lbw} considered other statistical ensembles, notably the canonical
ensemble in which the temperature and the mass are fixed.  In that case, the
equilibrium state is obtained by minimizing the Boltzmann free energy $F=E-TS$
at fixed mass $M$. The series of equilibria $T({\cal R})$
displays damped oscillations. No equilibrium state exists with a temperature
$T<T_c=0.397\, GMm/(k_BR)$, where $T_c$ corresponds to the first turning point
of temperature (with a density contrast
${\cal
R}'_c=32.1$).\footnote{These
results were first found by Emden.\cite{emden}} Using the Poincar\'e turning
point criterion, \cite{poincare} they concluded that equilibrium states
with a
density contrast ${\cal R}<32.1$ are thermodynamically stable (free energy
minima)
while equilibrium states with a density contrast ${\cal R}>32.1$ are
thermodynamically unstable (saddle points of free energy). Below $T_c$ the
system undergoes an ``isothermal collapse'' leading to a Dirac peak containing
all the particles. Since the stability
limits in the microcanonical and canonical ensembles differ, Lynden-Bell and
Wood \cite{lbw} encountered for the first time in statistical mechanics a
situation of
ensemble inequivalence. This is a peculiarity of systems with long-range
interactions.\cite{campabook} In the present context, it is related to the
fact
that
negative specific heats are allowed in the microcanonical ensemble while they
are forbidden in the canonical ensemble.\cite{lbw} Similar results were
obtained independently by Thirring.\cite{thirring}
Katz \cite{katzpoincare1} plotted the caloric curve $\beta(E)$ of
isothermal
self-gravitating spheres and exhibited its spiral behavior (see
Fig. \ref{elr}-a).\footnote{This spiral
behavior  was previously observed for
self-gravitating isothermal stars in other representations.
\cite{chandrabook,isouni1,henrich,schonberg,ebert,bonnor,crea} In the present
case, it is
associated
with the damped oscillations of energy and temperature as a function of the
density contrast.} He also extended the Poincar\'e theory on linear series of
equilibria\cite{poincare} to the case where there are several turning points and
developed a
general method to determine the thermodynamical stability of the equilibrium
states from the topology of the caloric curve $\beta(E)$. A
change of stability can only occur at a turning point of energy in the
microcanonical ensemble  or at a
turning point of temperature in the canonical ensemble. A mode of stability is
lost if the curve $\beta(-E)$ turns clockwise and gained if it turns
anticlockwise. In
this manner, one can determine the thermodynamical stability of 
the system by simply plotting the caloric curve (series of equilibria). The
seminal works of Antonov,\cite{antonov} Lynden-Bell and Wood,\cite{lbw} and
Katz\cite{katzpoincare1}
 were followed by many other studies (see, e.g., Refs.
\cite{paddyapj,dvsc,dvs1,dvs2,aaiso,grand,katzokamoto,metastable}).

\def\figsubcap#1{\par\noindent\centering\footnotesize(#1)}
\begin{figure}[h]%
\begin{center}
\parbox{2.1in}{\includegraphics[width=2in]{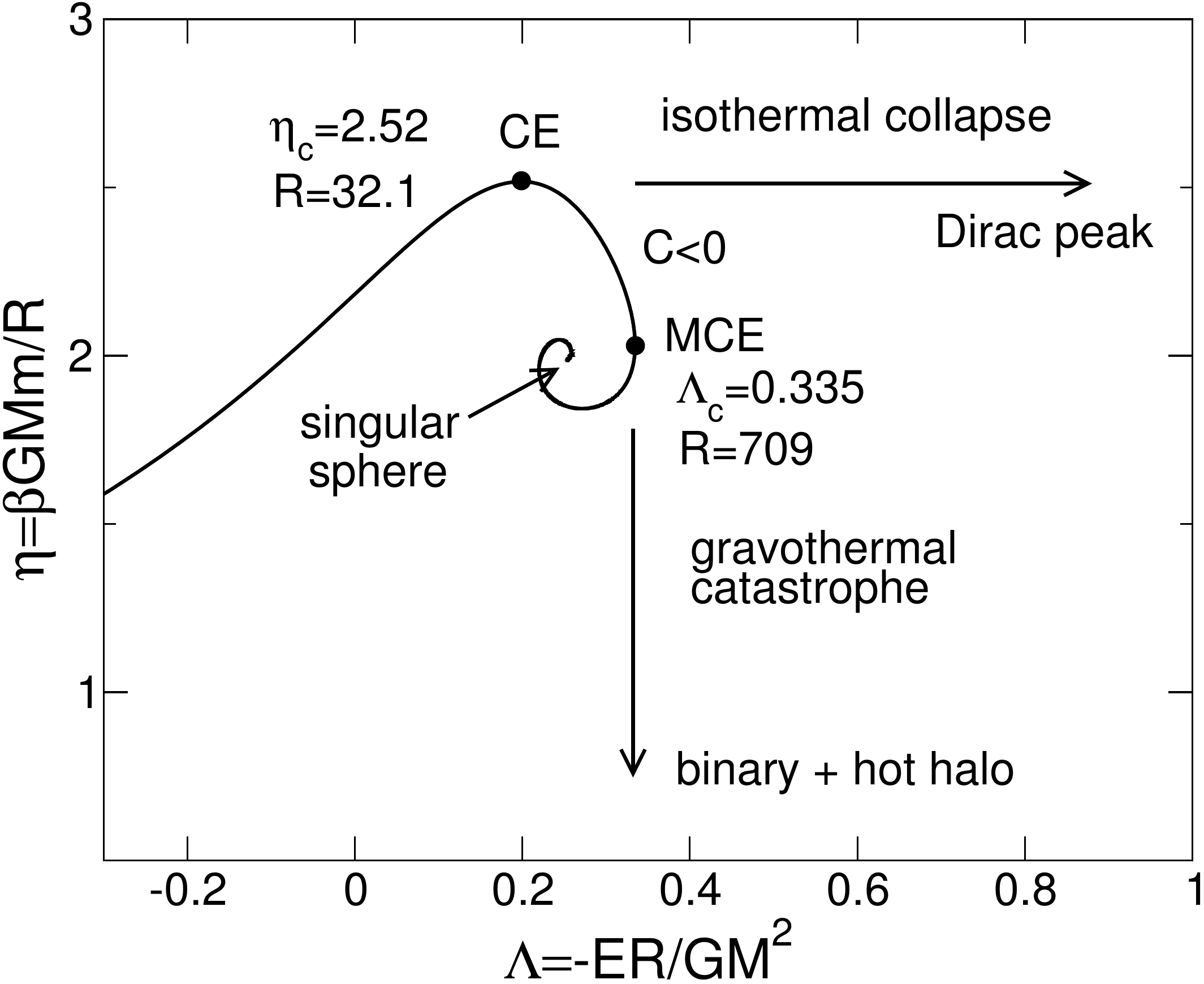}\figsubcap
{a}}
  \hspace*{4pt}
  \parbox{2.1in}{\includegraphics[width=2in]{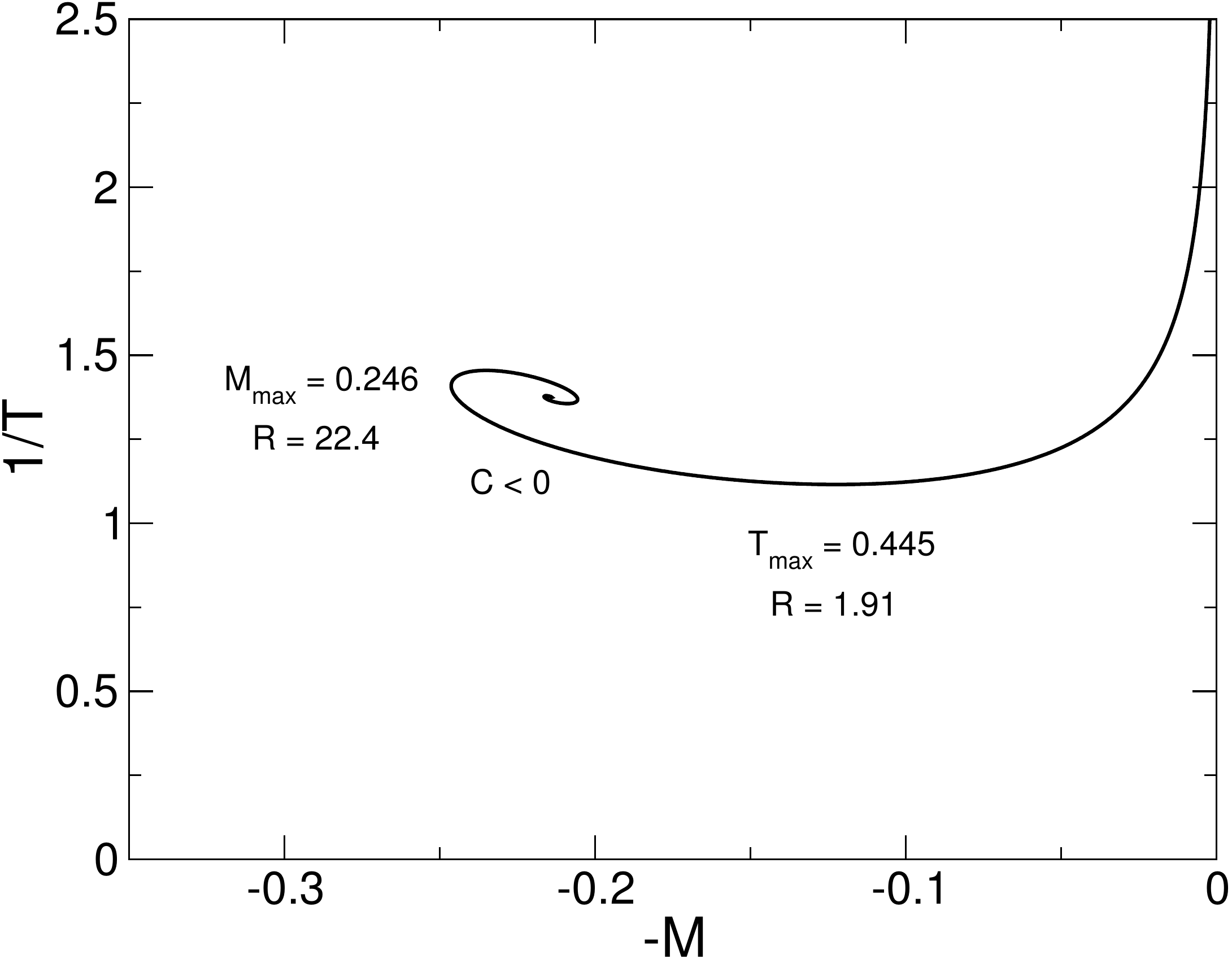}\figsubcap{b}}
  \caption{(a) Caloric curve
of nonrelativistic classical self-gravitating 
particles (cold spiral).\cite{ijmpb} We have plotted the normalized inverse
temperature
$\eta=\beta GMm/R$ as a function of minus the
normalized binding energy $\Lambda=-ER/GM^2$. (b) Caloric curve of the
self-gravitating radiation in general relativity (hot
spiral).\cite{aarelat2} We have plotted the normalized inverse temperature
$\hbar^{3/4}c^{7/4}/(k_B T_{\infty} G^{1/4}R^{1/2})$ as a function
of minus the
normalized energy $-G M/Rc^2$.}
  \label{elr}
\end{center}
\end{figure}

The statistical mechanics of the self-gravitating black-body
radiation (photon star) confined within a cavity in general
relativity was investigated by Sorkin {\it et al.}\cite{sorkin} and, more
recently, by Chavanis.\cite{aarelat2} They showed that the caloric curve
$\beta_{\infty}({\cal E})$ forms a spiral (see
Fig. \ref{elr}-b). There is no equilibrium state above a
maximum mass-energy $M_{\rm max}c^2=0.246\, Rc^4/G$ (corresponding to a
density contrast $22.4$) or above a
maximum temperature $(T_{\infty})_{\rm
max}=0.445\, \hbar^{3/4}c^{7/4}/(k_BG^{1/4}R^{1/2})$ 
(corresponding to a density contrast $1.91$).
In that case, the
system is expected
to collapse towards a black hole. We note that the ``hot spiral'' (see Fig.
\ref{elr}-b) of the
self-gravitating radiation in general relativity (ultrarelativistic limit) is
inverted with respect to
the ``cold spiral'' of the nonrelativistic classical self-gravitating gas (see
Fig.
\ref{elr}-a).

\section{Classical particles at $T>0$ in general relativity}
\label{sec_rc}

The statistical mechanis of classical particles in general relativity has been
considered by Roupas\cite{roupas} and, independently, by Alberti and
Chavanis.\cite{acb} The caloric curve depends on one parameter, the particle
number $N$ (more precisely $N/R$).
Generically, the caloric curve $\beta_{\infty}(E)$ has the form of a
double spiral (see Fig. \ref{rcs}) which combines the aspects of the ``cold
spiral''
corresponding to a
nonrelativistic gas and the aspects of the ``hot spiral'' corresponding to an
ultrarelativistic gas discussed in Sec. \ref{sec_nrc}.\footnote{The hot spiral
of
ultrarelativistic classical particles in general relativity is similar, but not
identical, to the hot spiral of the  self-gravitating
radiation.\cite{aarelat2,acb,gr1}} There is no equilibrium state below a minimum
energy (resp. minimum
temperature) and above a maximum energy (resp. maximum temperature) in the
microcanonical (resp. canonical) ensemble. When the number of particles $N$
increases, the two spirals approach each other, merge, form a loop, and
finally
disappear (by reducing to a point) at $N_{\rm max}=0.1764\, Rc^2/Gm$. For
$N>N_{\rm max}$, there is no
equilibrium state whatever the value of mass-energy and temperature.

\def\figsubcap#1{\par\noindent\centering\footnotesize(#1)}
\begin{figure}[h]%
\begin{center}
\parbox{2.1in}{\includegraphics[width=2in]{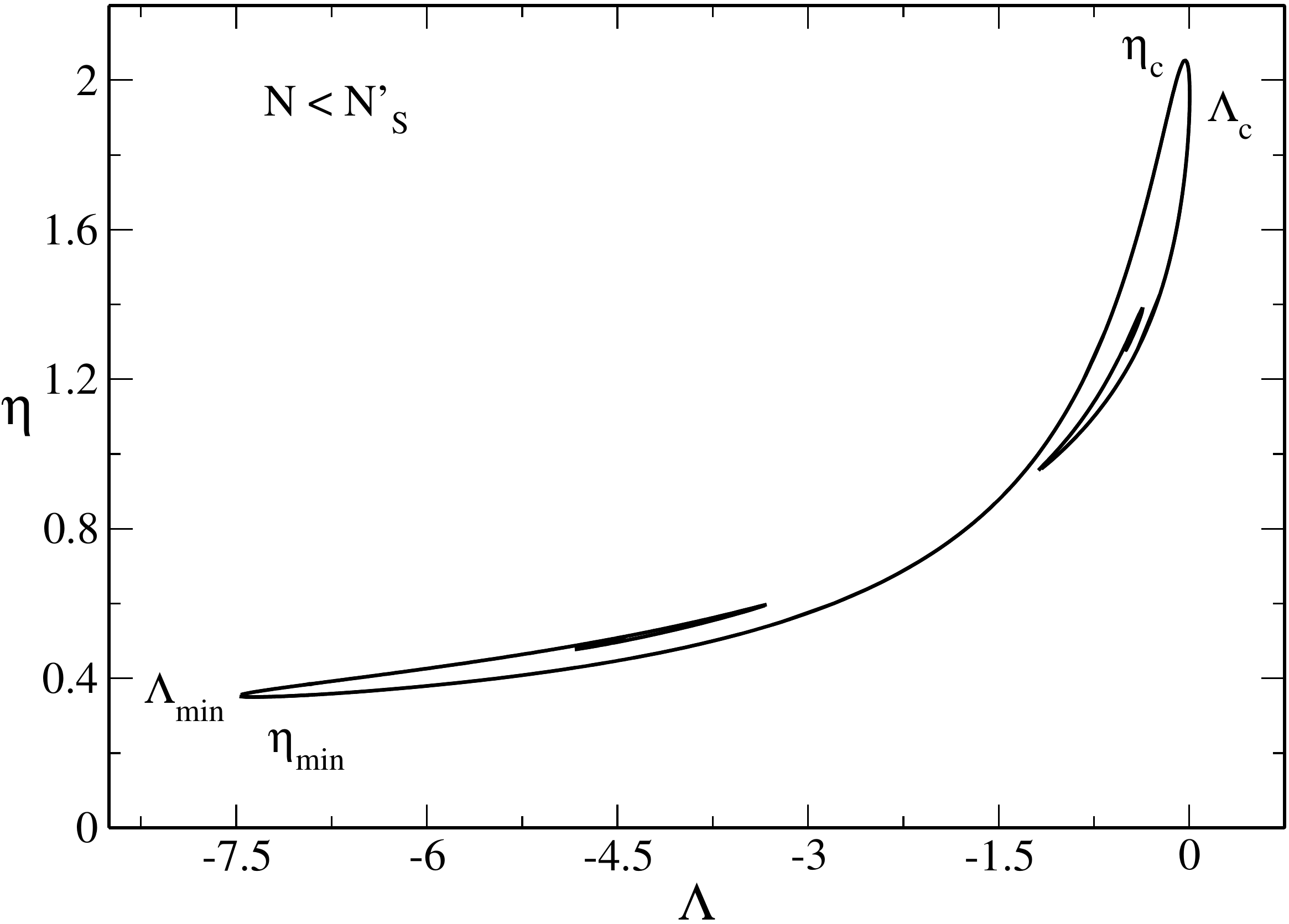}\figsubcap
{a}}
  \hspace*{4pt}
  \parbox{2.1in}{\includegraphics[width=2in]{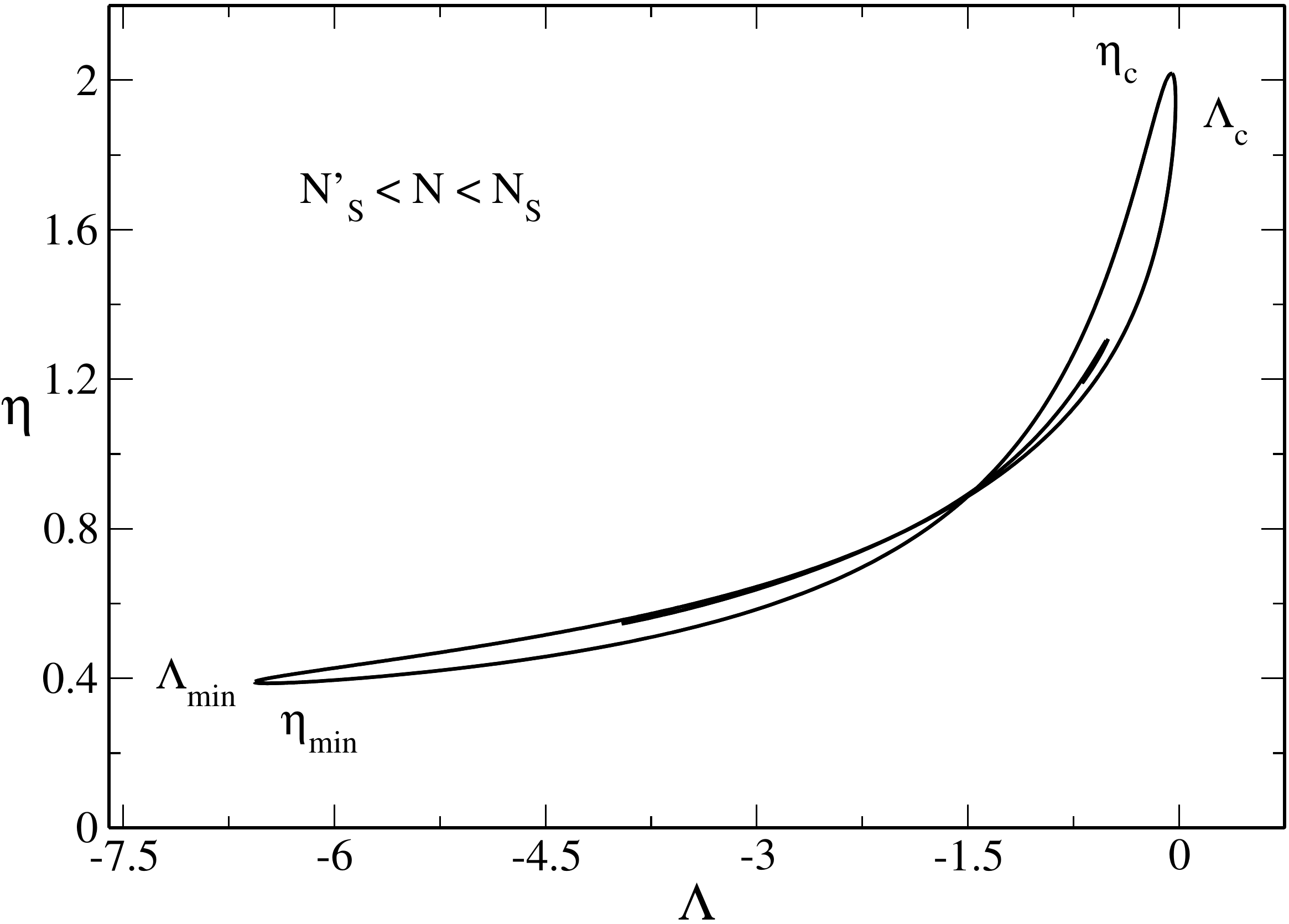}\figsubcap{b}}
\parbox{2.1in}{\includegraphics[width=2in]{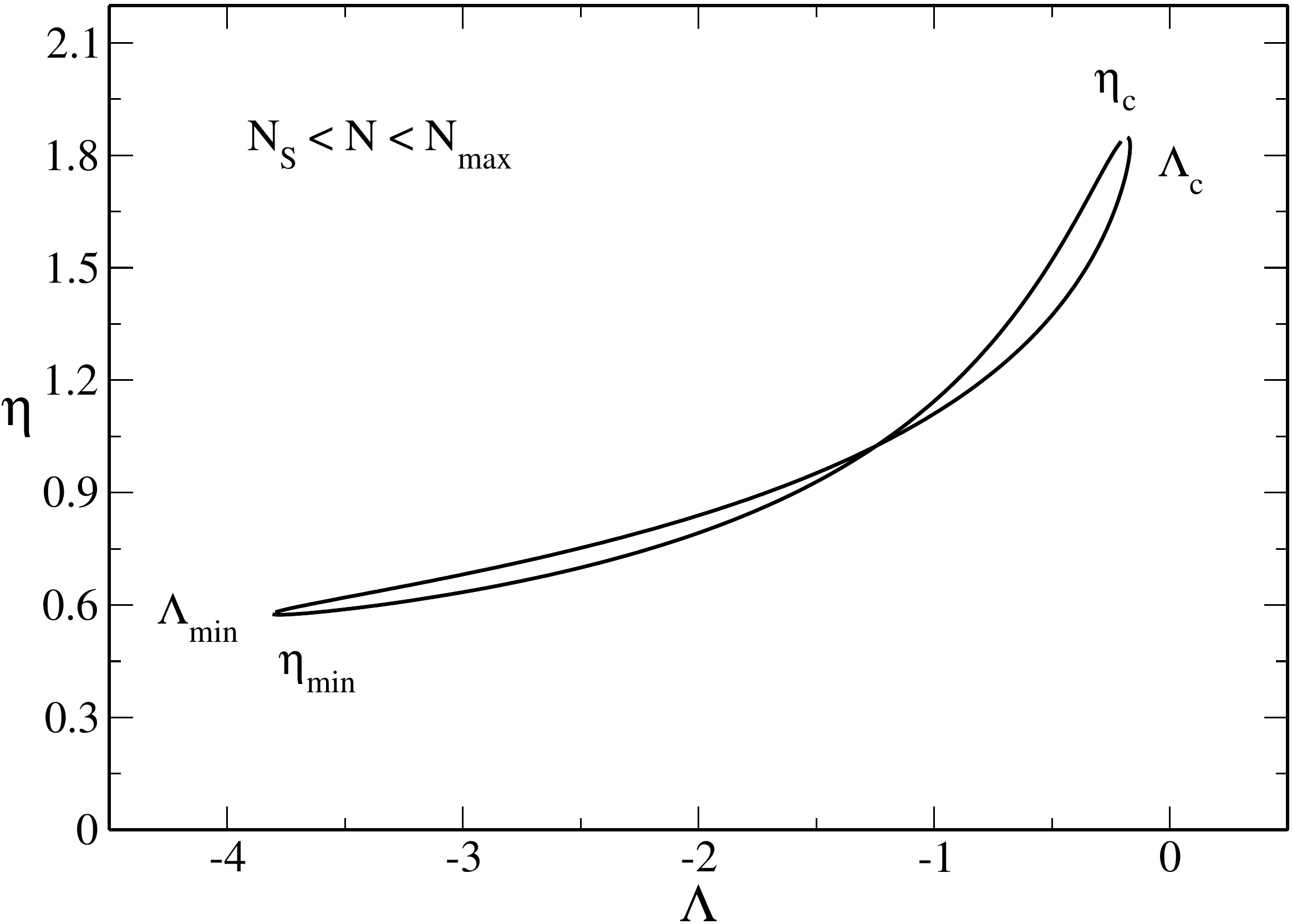}\figsubcap{c}}
\parbox{2.1in}{\includegraphics[width=2in]{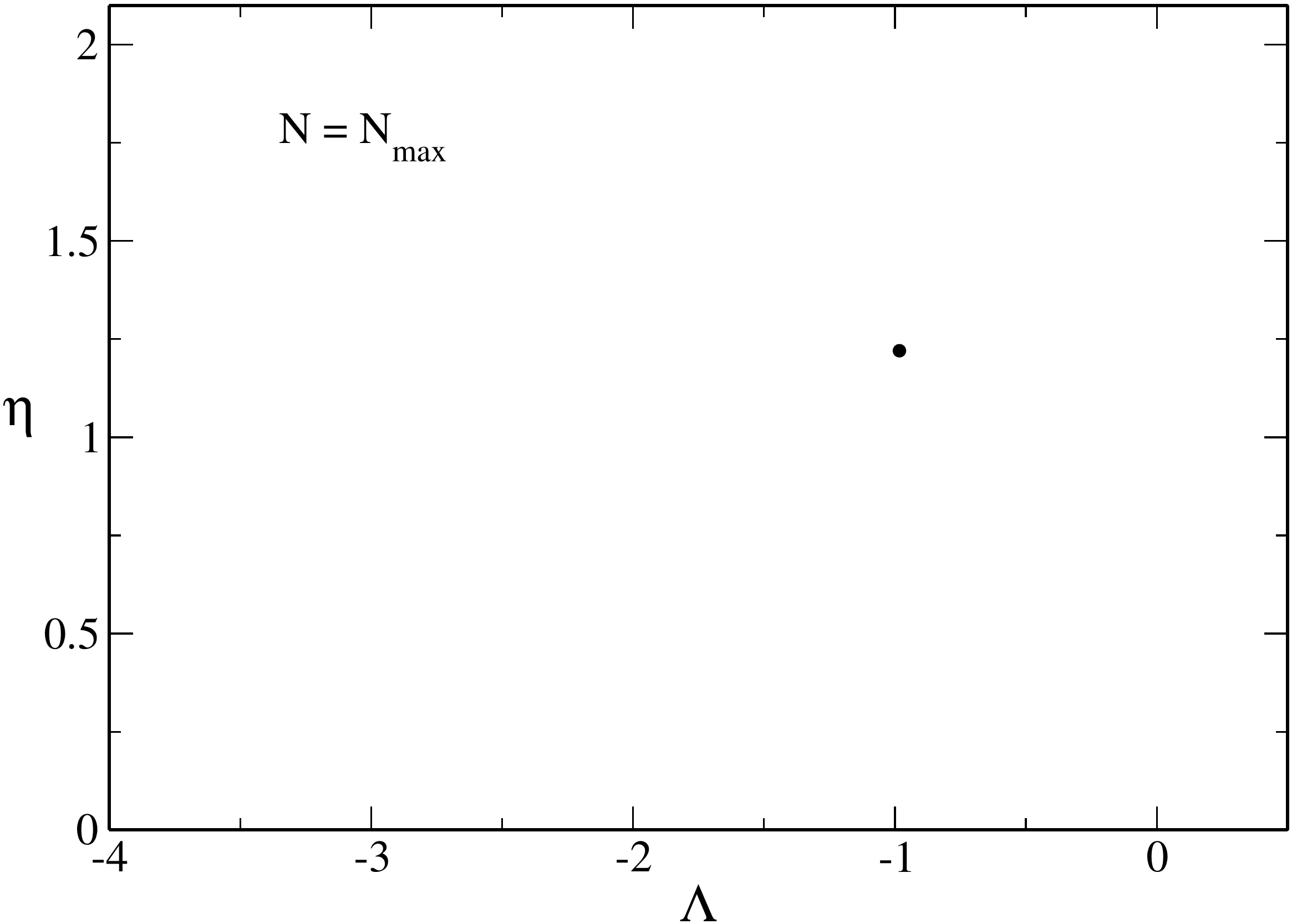}\figsubcap{d}}
  \caption{Caloric curve of classical particles in general relativity.
We
have plotted the normalized inverse temperature $\eta=\beta_{\infty} GNm^2/R$ as
a function of minus
the normalized binding energy $\Lambda=-ER/GN^2m^2$ for different values of the
normalized
particle number $\nu=GNm/Rc^2$. (a) Double spiral
(b) Merging (c) Loop (d) Point.\cite{acb}}%
  \label{rcs}
\end{center}
\end{figure}

\section{Nonrelativistic fermions at $T>0$}
\label{sec_nrf}

The statistical mechanics of self-gravitating fermions at
nonzero 
temperature was first studied by Hertel and Thirring \cite{htf,ht} as a simple
(nonrelativistic) model of neutron stars. In that case, the density profile
decreases  with the distance as $r^{-2}$ and we need to confine the
system within a box  in order to have an
equilibrium state with a finite mass. Hertel and Thirring \cite{htf} rigorously
proved that the mean field approximation (which amounts to neglecting
correlations among the particles) and the Thomas-Fermi approximation (which
amounts to neglecting the quantum potential) become exact in a proper
thermodynamic limit where $N\rightarrow +\infty$. This leads to the
Fermi-Dirac-Poisson equations, also
known as the temperature-dependent Thomas-Fermi equations. Hertel and Thirring
\cite{ht} solved these equations numerically and plotted the caloric curve.
The caloric curve depends on one parameter, the box radius $R$ (more precisely
$NR^3$). They found
that, for
sufficiently
large systems ($R>R_{\rm CCP}=12.8\, \hbar^2/(N^{1/3}Gm^3)$), a
negative specific heat region occurs in the
microcanonical
ensemble (see Fig. \ref{rfc}). Since negative specific heats are forbidden in
the canonical ensemble,
this implies that the statistical ensembles are inequivalent. The region of
negative specific heats which is allowed in the microcanonical ensemble is
replaced by a first order phase transition in the canonical ensemble. This phase
transition is expected to take place at a transition temperature $T_t$
connecting the gaseous phase to the condensed phase through a horizontal Maxwell
plateau in the caloric curve $T(E)$. This is accompanied by a 
discontinuity of energy. There is also a lower critical temperature $T_c$
(spinodal point) at
which the metastable gaseous phase disappears and the system  collapses (zeroth
order phase transition). This collapse is associated with the isothermal
collapse of classical self-gravitating systems.\cite{emden,aaiso} However, for
self-gravitating fermions, the collapse stops
when quantum degeneracy comes into play. In that case, the system
achieves a ``core-halo'' configuration made of a quantum core (fermion ball)
containing almost all the mass surrounded by a tenuous isothermal atmosphere.
There is also a higher critical temperature $T_*$ (spinodal point) at which the
metastable condensed phase disappears and the system explodes. These results
were exported by Bilic and Viollier \cite{bvn} to the context of dark matter.

\def\figsubcap#1{\par\noindent\centering\footnotesize(#1)}
\begin{figure}[h]%
\begin{center}
\parbox{2.1in}{\includegraphics[width=2in]{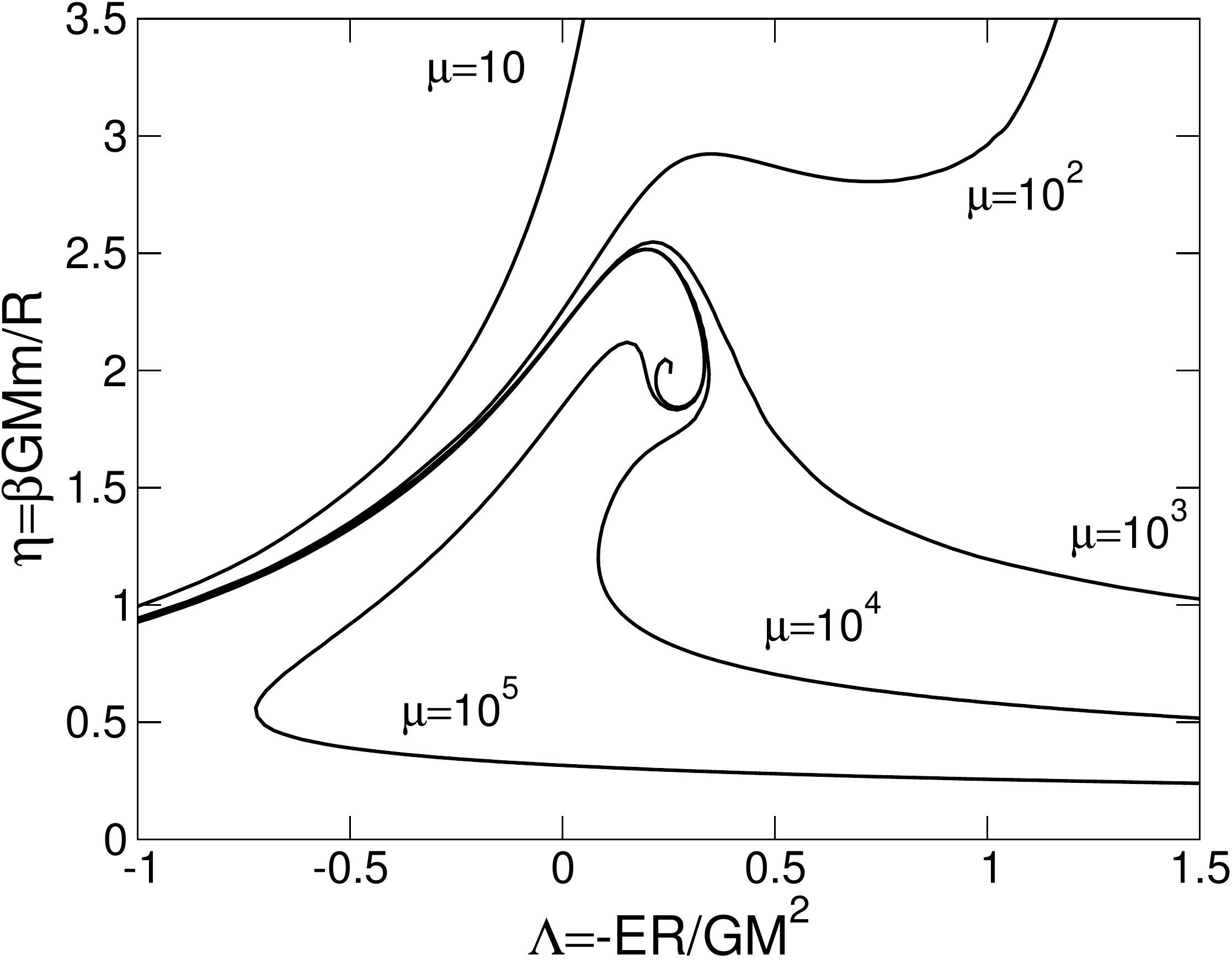}
\figsubcap{a}}
  \hspace*{4pt}
\parbox{2.1in}{\includegraphics[width=2in]{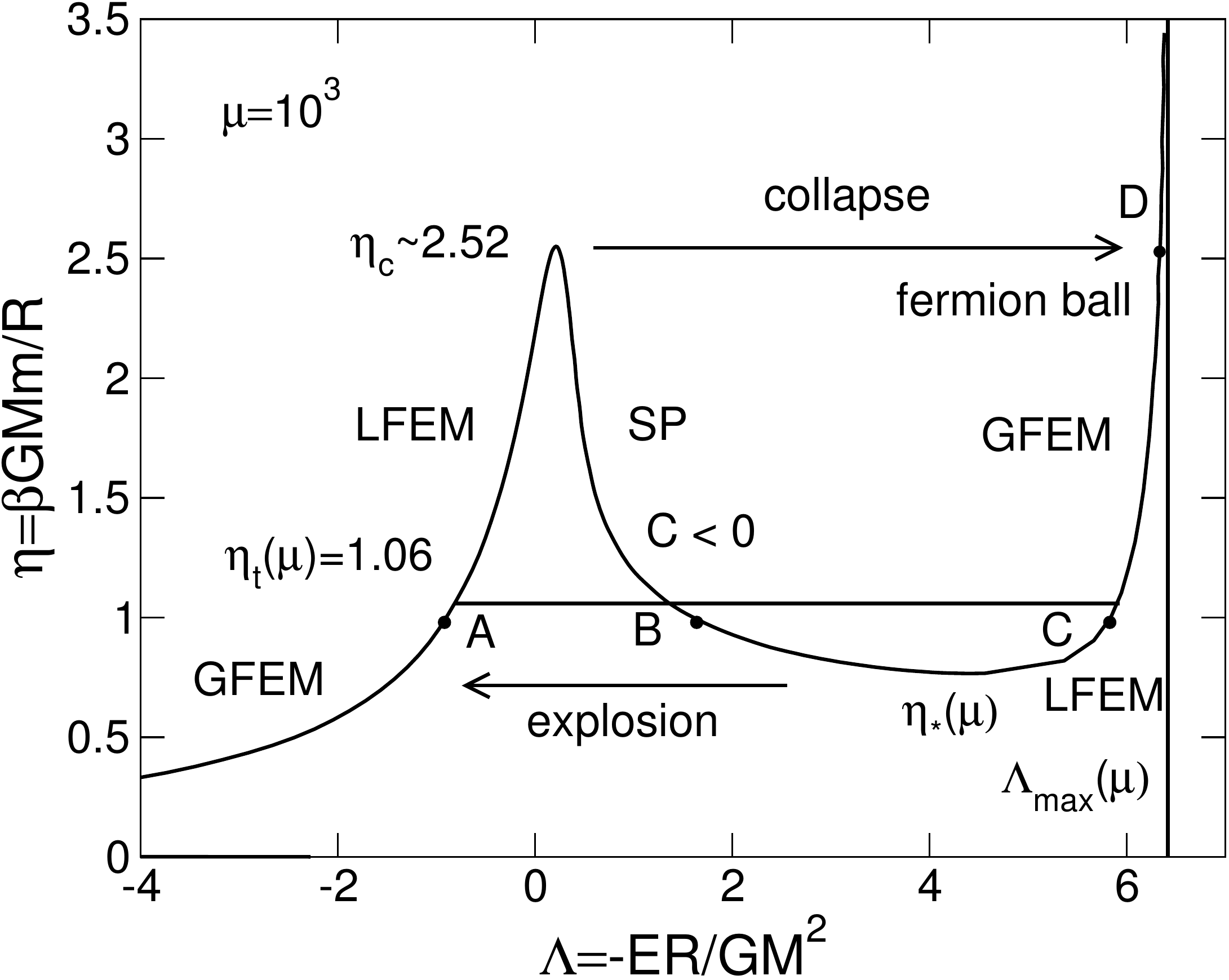}
\figsubcap{b}}
  \caption{Caloric curve of nonrelativistic
self-gravitating fermions.\cite{ijmpb} We
have plotted the normalized inverse temperature $\eta=\beta GMm/R$ as a function
of minus
the normalized energy $\Lambda=-ER/GM^2$ for different values of the normalized
radius, or degeneracy
parameter $\mu=\eta_0\sqrt{512\pi^4G^3MR^3}$,  where
$\eta_0={2m^4}/{h^3}$ is the maximum possible value of the distribution
function fixed by the Pauli exclusion principle. (a) Dependence of the series of
equilibria on the size of the system. (b)
For small systems, the caloric curve has an $N$-shape
structure.}
  \label{rfc}
\end{center}
\end{figure}

\def\figsubcap#1{\par\noindent\centering\footnotesize(#1)}
\begin{figure}[h]%
\begin{center}
\parbox{2.1in}{\includegraphics[width=2in]{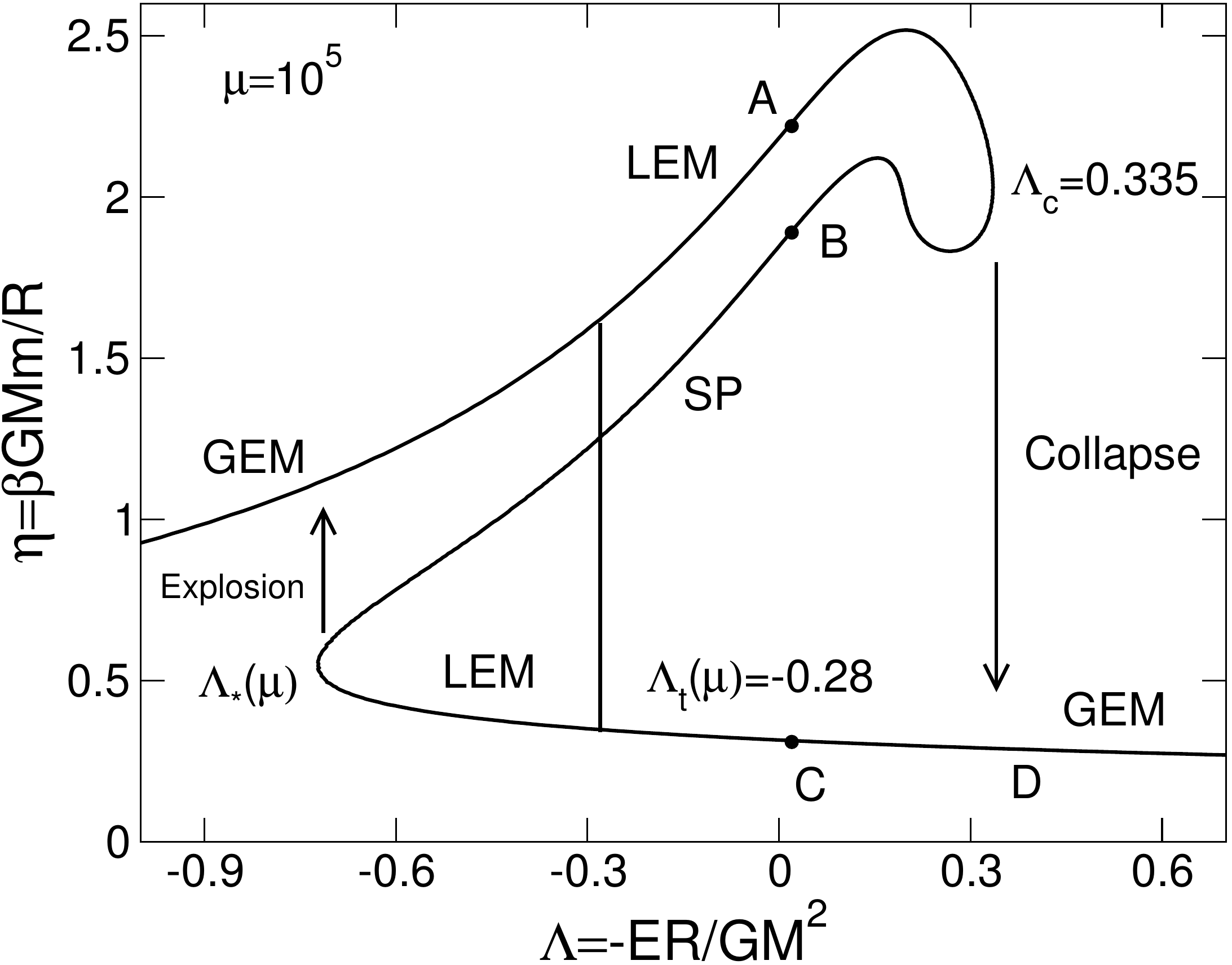}
\figsubcap{a}}
  \hspace*{30pt}
\parbox{2.1in}{\includegraphics[width=2in]{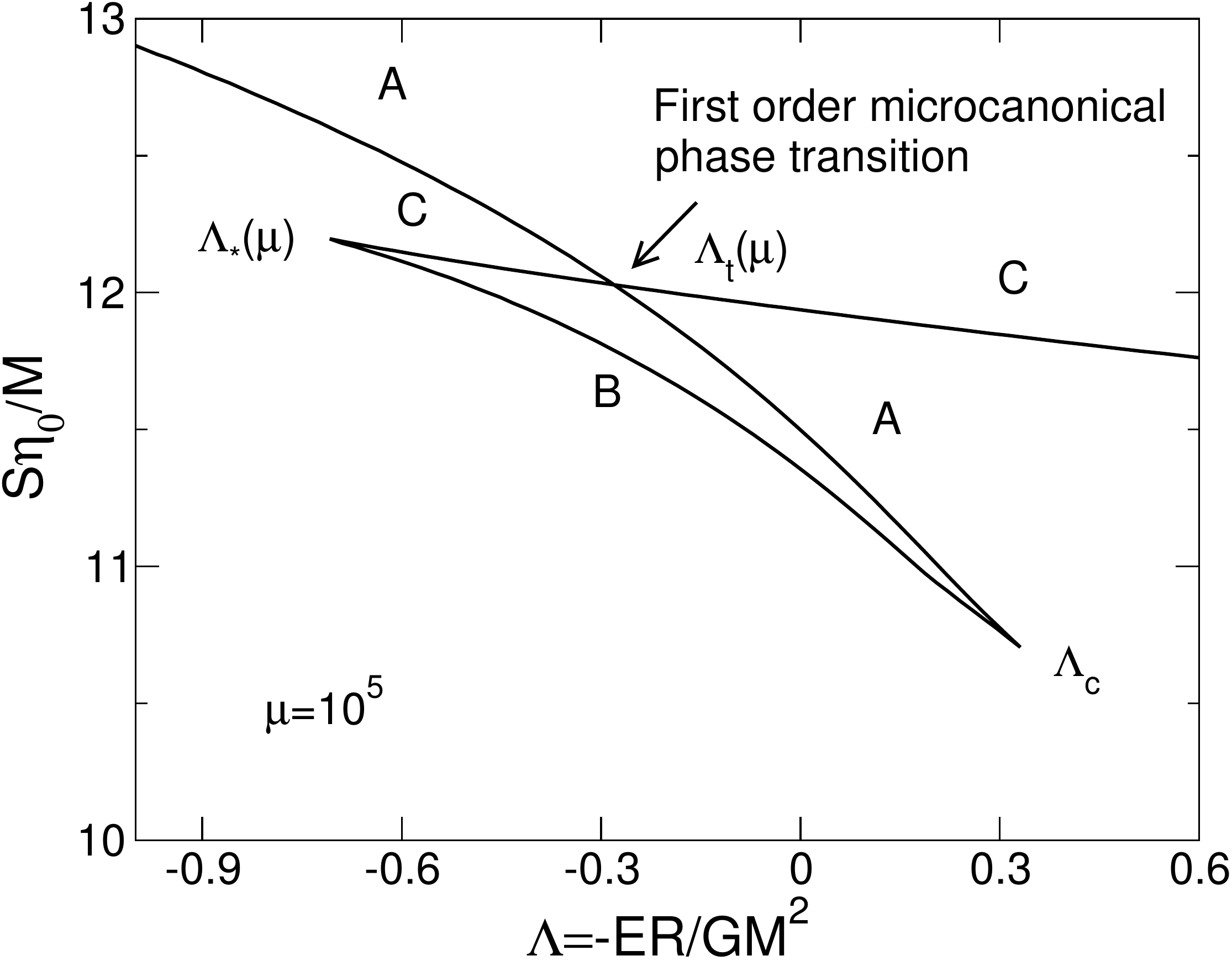}
\figsubcap{b}}
  \caption{Caloric curve of nonrelativistic self-gravitating
fermions.\cite{ijmpb} (a)
For large systems, the caloric curve has a $Z$-shape structure
resembling a dinosaur's
neck. (b) Illustration of the microcanonical first order phase
transition on the $S(E)$ curve.
}
  \label{rfm}
\end{center}
\end{figure}

\def\figsubcap#1{\par\noindent\centering\footnotesize(#1)}
\begin{figure}[h]%
\begin{center}
\parbox{2.1in}{\includegraphics[width=2in]{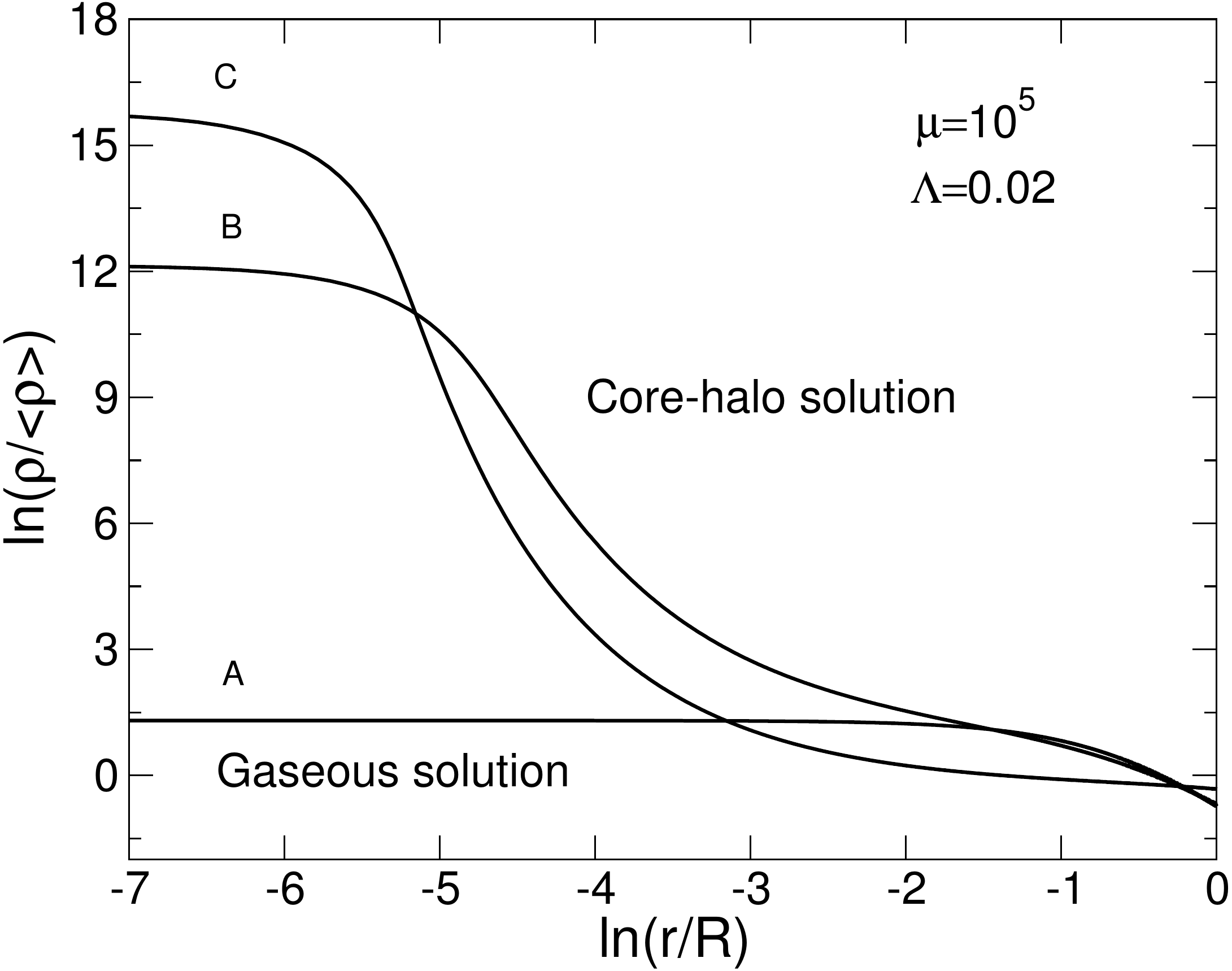}
\figsubcap{a}}
  \hspace*{30pt}
\parbox{2.1in}{\includegraphics[width=2in]{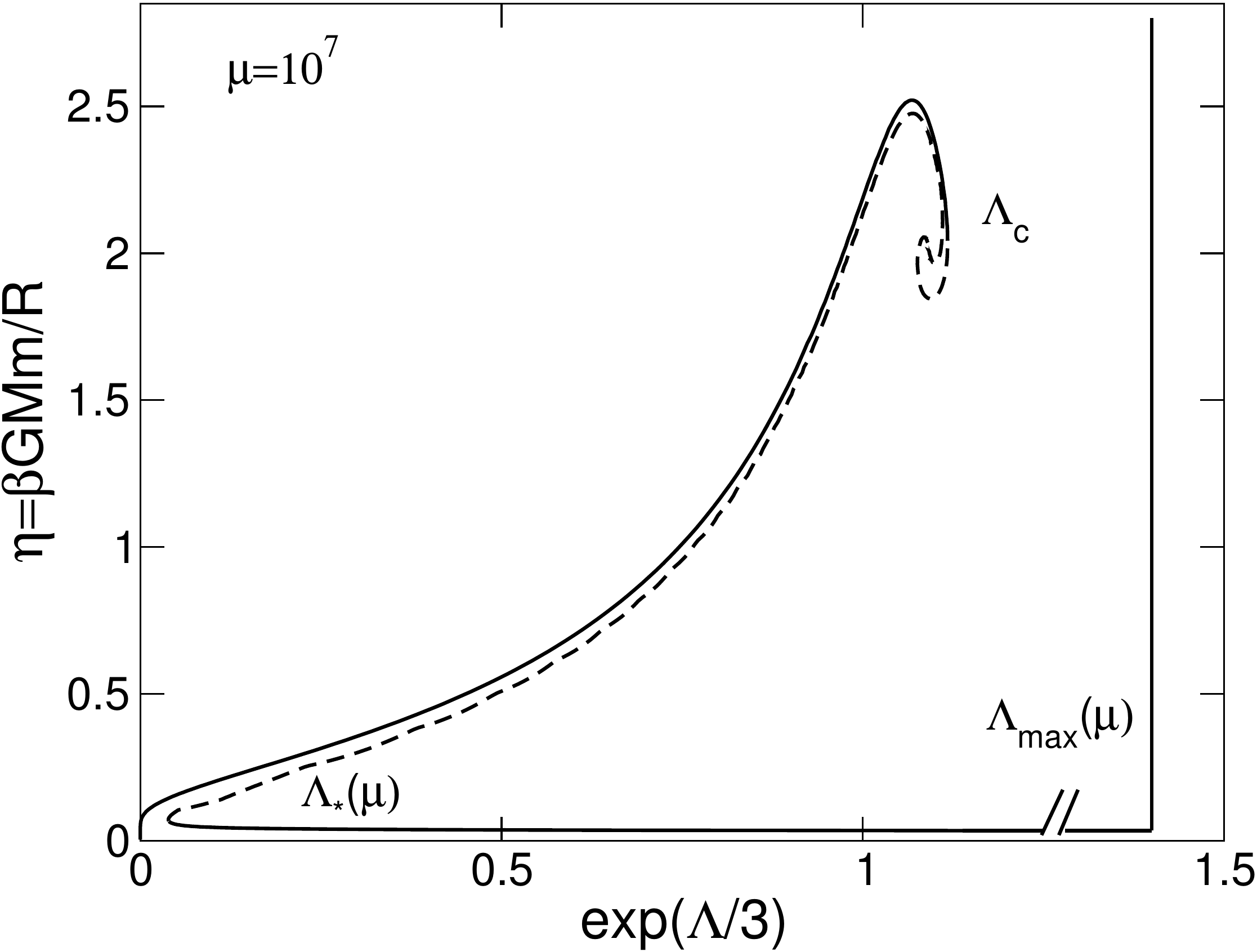}
\figsubcap{b}}
  \caption{(a) Density profiles corresponding to the gaseous and
core-halo solutions.\cite{ijmpb} (b) The classical limit $\mu\rightarrow
+\infty$ (very large systems).\cite{ijmpb} According to the Poincar\'e-Katz
criterion,\cite{poincare,katzpoincare1} the equilibrium states are unstable in
the
microcanonical (resp. canonical) ensemble between the first and the last
turning points of energy (resp. temperature).}
  \label{dp}
\end{center}
\end{figure}

An exhaustive study of phase transitions in the self-gravitating 
Fermi gas was made by Chavanis \cite{pt,ijmpb} in both canonical and
microcanonical ensembles. He confirmed the phase transition in the canonical
ensemble previously found by Hertel and Thirring \cite{ht} and evidenced, for
sufficiently large
systems ($R>R_{\rm MCP}=130\, \hbar^2/(N^{1/3}Gm^3)$), a first
order phase
transition in the microcanonical ensemble (see Fig. \ref{rfm}).
This
phase transition is expected to take place at a transition energy $E_t$
connecting the gaseous phase to the condensed phase through a vertical Maxwell
plateau in the caloric curve $T(E)$. This is accompanied by a 
discontinuity of temperature. There is also a lower
critical energy $E_c$ (spinodal point) at
which the metastable gaseous phase disappears and the system collapses (zeroth
order phase
transition). This collapse is associated with the gravothermal catastrophe of
classical self-gravitating systems.\cite{antonov,lbw} However,  for
self-gravitating fermions, the collapse stops when quantum
degeneracy comes into play. In that case, the system achieves a
core-halo configuration\cite{csmnras} made of a quantum core (fermion ball)
containing a
moderate fraction of the total mass surrounded by a massive isothermal
atmosphere (see Fig. \ref{dp}-a). There is also a higher critical
energy $E_*$ (spinodal
point) at
which the metastable condensed phase disappears and the system explodes.  As a
result, there exist two distinct critical points in the self-gravitating Fermi
gas, one in each ensemble, above which phase transitions occur. For small
systems ($R<R_{\rm CCP}$), there
is no phase transition, for intermediate size systems ($R_{\rm CCP}<R<R_{\rm
MCP}$) a phase transition takes place
in the canonical ensemble but not in the microcanoncal ensemble, and for large
systems ($R>R_{\rm MCP}$) a phase transition takes place in both ensembles.
When $R\rightarrow +\infty$, the series of equilibria rotates several times
before unwinding (see Fig. \ref{dp}-b) and we recover the classical spiral from
Fig. \ref{elr}-a. Chavanis\cite{ijmpb}
determined the phase diagram of the nonrelativistic self-gravitating Fermi gas.
He also argued
that the lifetime of
metastable states is extremely long, scaling as $e^N$ where $N$ is the number of
particles, so that the first order phase transitions do not take place in
practice.\cite{lifetime} Only zeroth order phase transitions associated with
the spinodal points are physically meaningful.

\section{General relativistic fermions at $T>0$}

The statistical mechanics of self-gravitating fermions at
nonzero
temperature in 
general relativity was first considered by Bilic and Viollier.\cite{bvr} As
before, the system has to be confined within a spherical box of radius
$R$ in order 
to prevent its evaporation. The caloric curve depends on two parameters, the
box radius $R$ and the particle number $N$. Bilic and
Viollier \cite{bvr} studied the case where
$R$ is relatively
small and $N<N_{\rm OV}$. In that case, the results are qualitatively similar to
those obtained for nonrelativistic fermions (see Sec. \ref{sec_nrf}). There is a
first order phase transition in the canonical ensemble which replaces the region
of negative specific heat present in the microcanonical ensemble. An equilibrium
state, which  is either ``gaseous'' (corresponding to the classical isothermal
sphere) or ``condensed'' (made of a fermion ball
surrounded by a classical isothermal envelope), exists  for any value of
temperature $T_{\infty}$
and binding energy $E$. General relativistic effects only introduce a small
correction to the Newtonian results.

\def\figsubcap#1{\par\noindent\centering\footnotesize(#1)}
\begin{figure}[h]%
\begin{center}
\parbox{2.1in}{\includegraphics[width=2in]{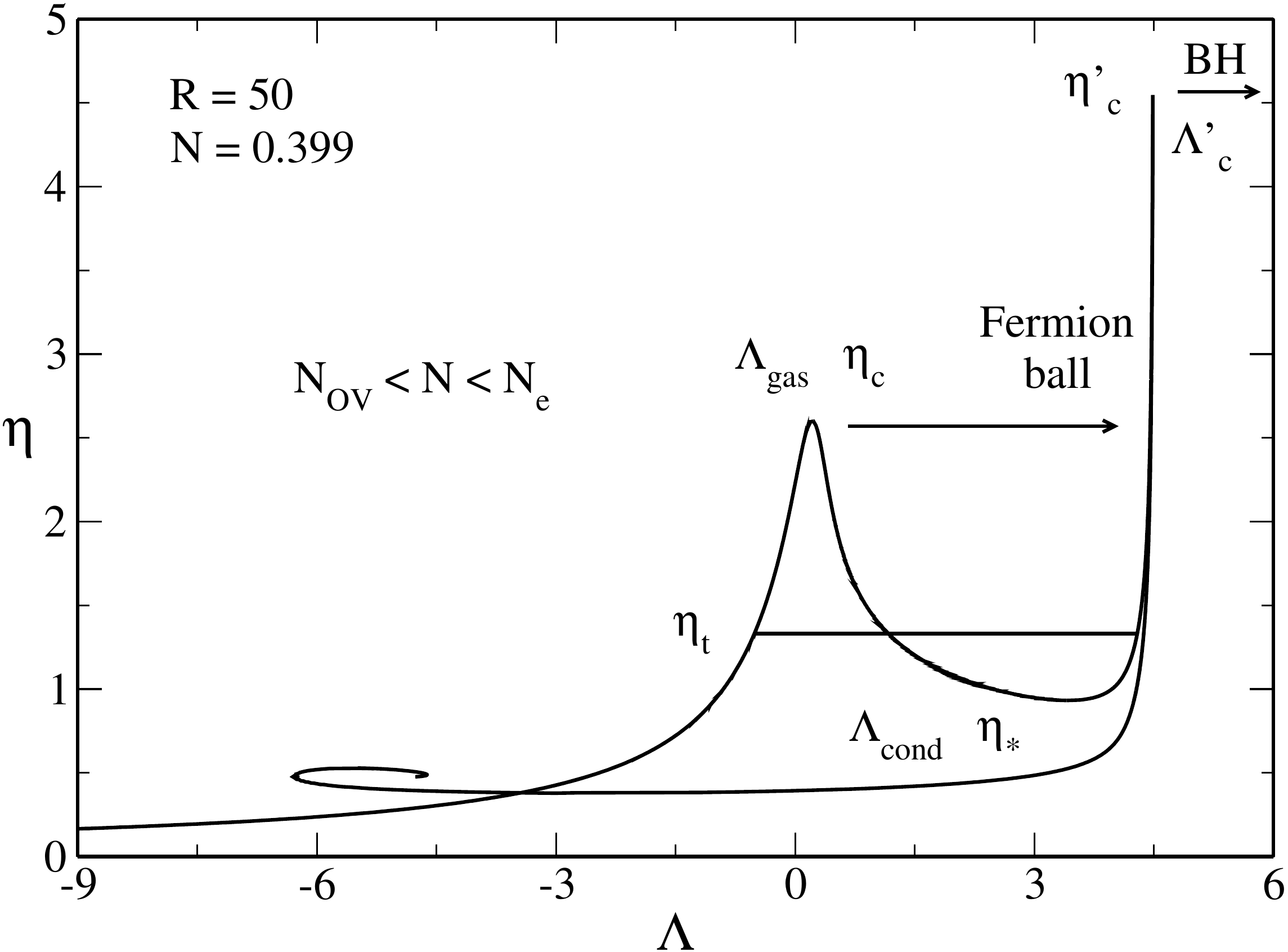}
\figsubcap{a}}
  \hspace*{4pt}
\parbox{2.1in}{\includegraphics[width=2in]{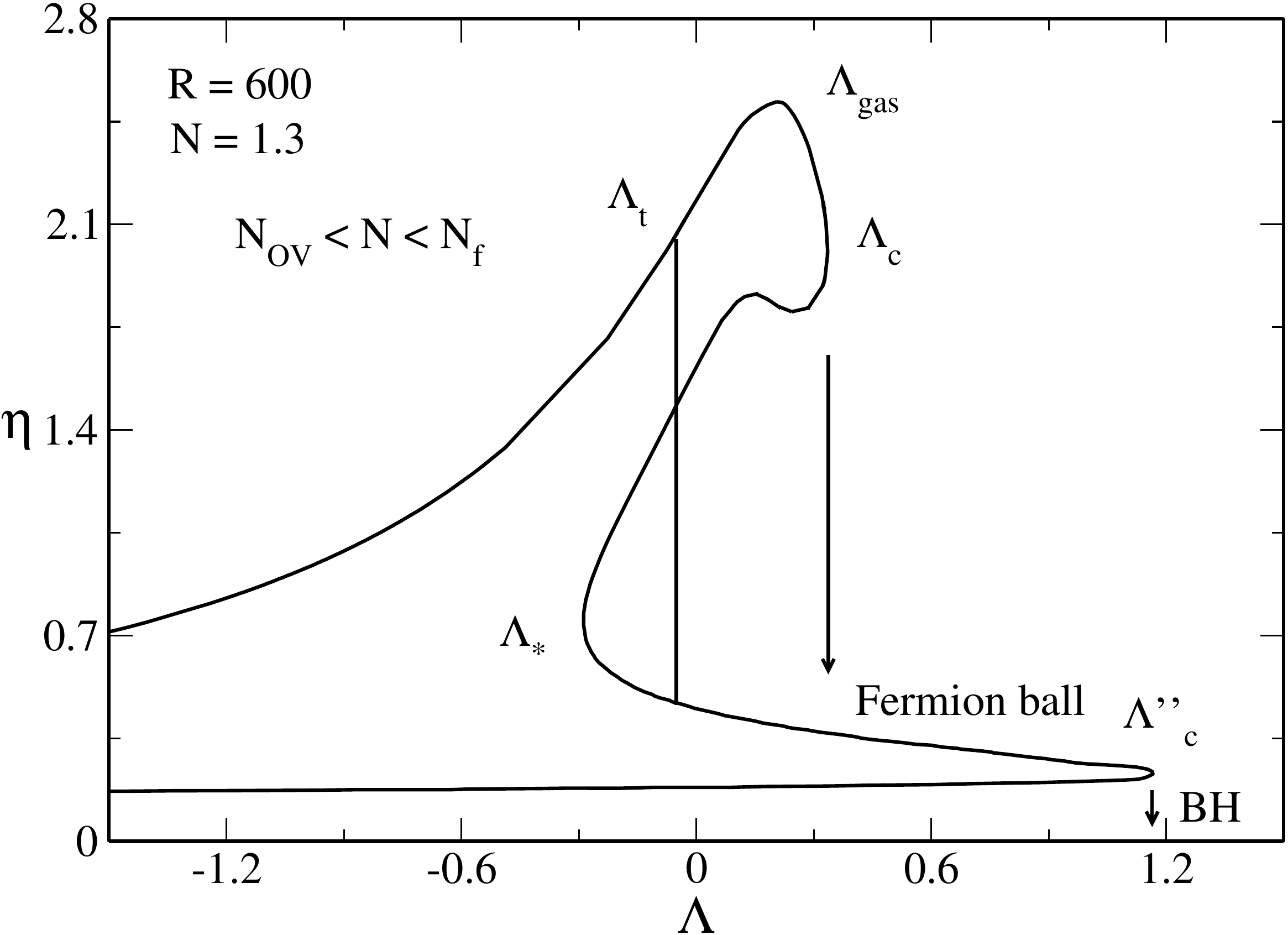}
\figsubcap{b}}
  \caption{Caloric curve of self-gravitating
fermions in general
relativity when $N>N_{\rm OV}$.\cite{acf} We
have plotted the normalized inverse temperature $\eta=\beta_{\infty} GNm^2/R$ as
a function
of minus
the normalized binding energy $\Lambda=-ER/GN^2m^2$ for different values of the
normalized
radius $R/R_{\rm OV}$ and normalized particle number $N/N_{\rm OV}$. (a)
Small systems: As $T$ decreases the system experiences an isothermal collapse
at $T_c$ leading to a fermion ball, then a gravitational collapse  at
$T'_c$ leading to a black hole. (b) Large systems: As $E$
decreases the system experiences a gravothermal catastrophe at $E_c$ leading to
a 
fermion ball surrounded by a hot halo, then a gravitational collapse  at
$E''_c$ leading to a black hole.}
  \label{ac1}
\end{center}
\end{figure}

\def\figsubcap#1{\par\noindent\centering\footnotesize(#1)}
\begin{figure}[h]%
\begin{center}
\parbox{2.1in}{\includegraphics[width=2in]{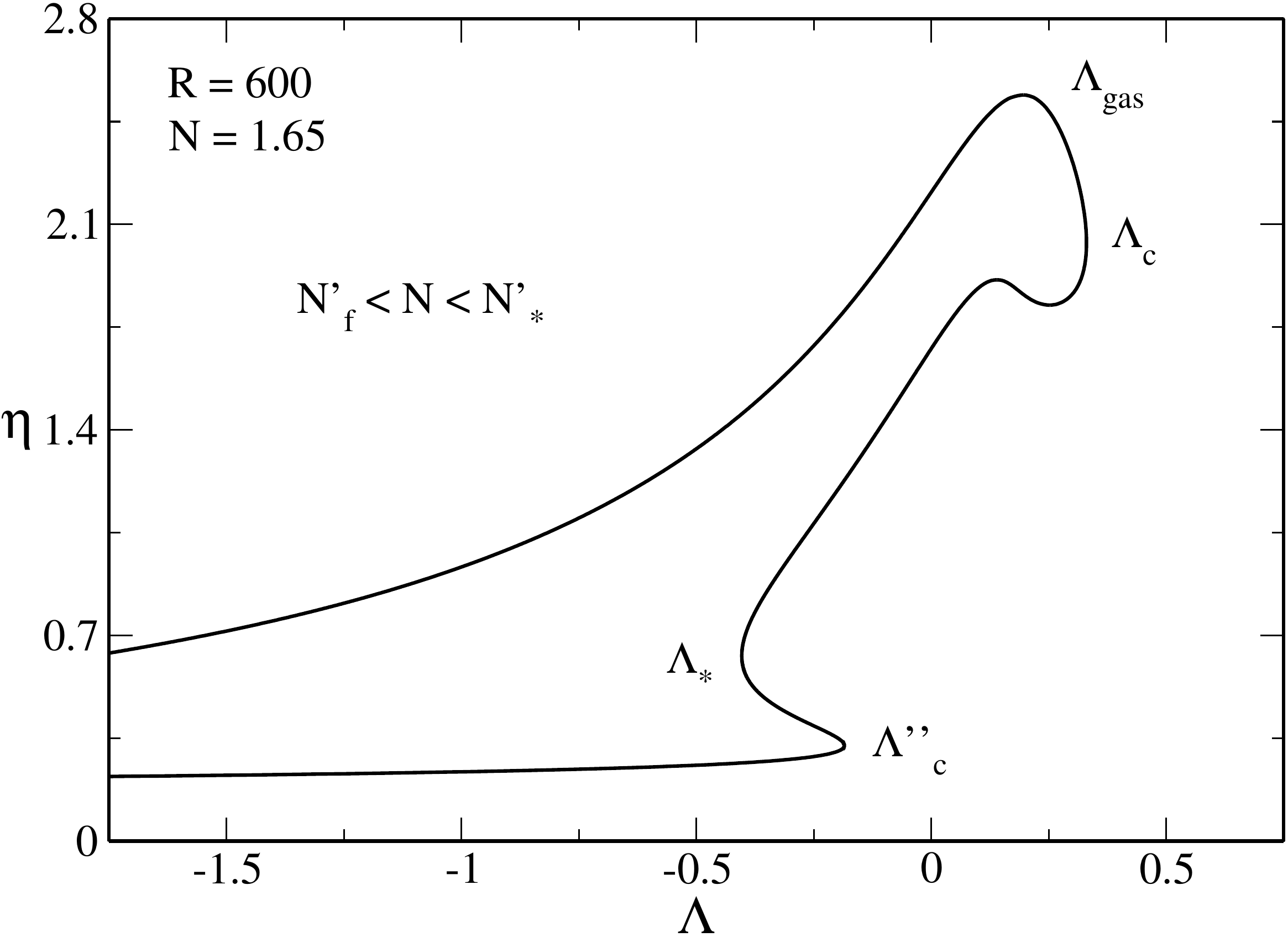}
\figsubcap{a}}
  \hspace*{4pt}
\parbox{2.1in}{\includegraphics[width=2in]{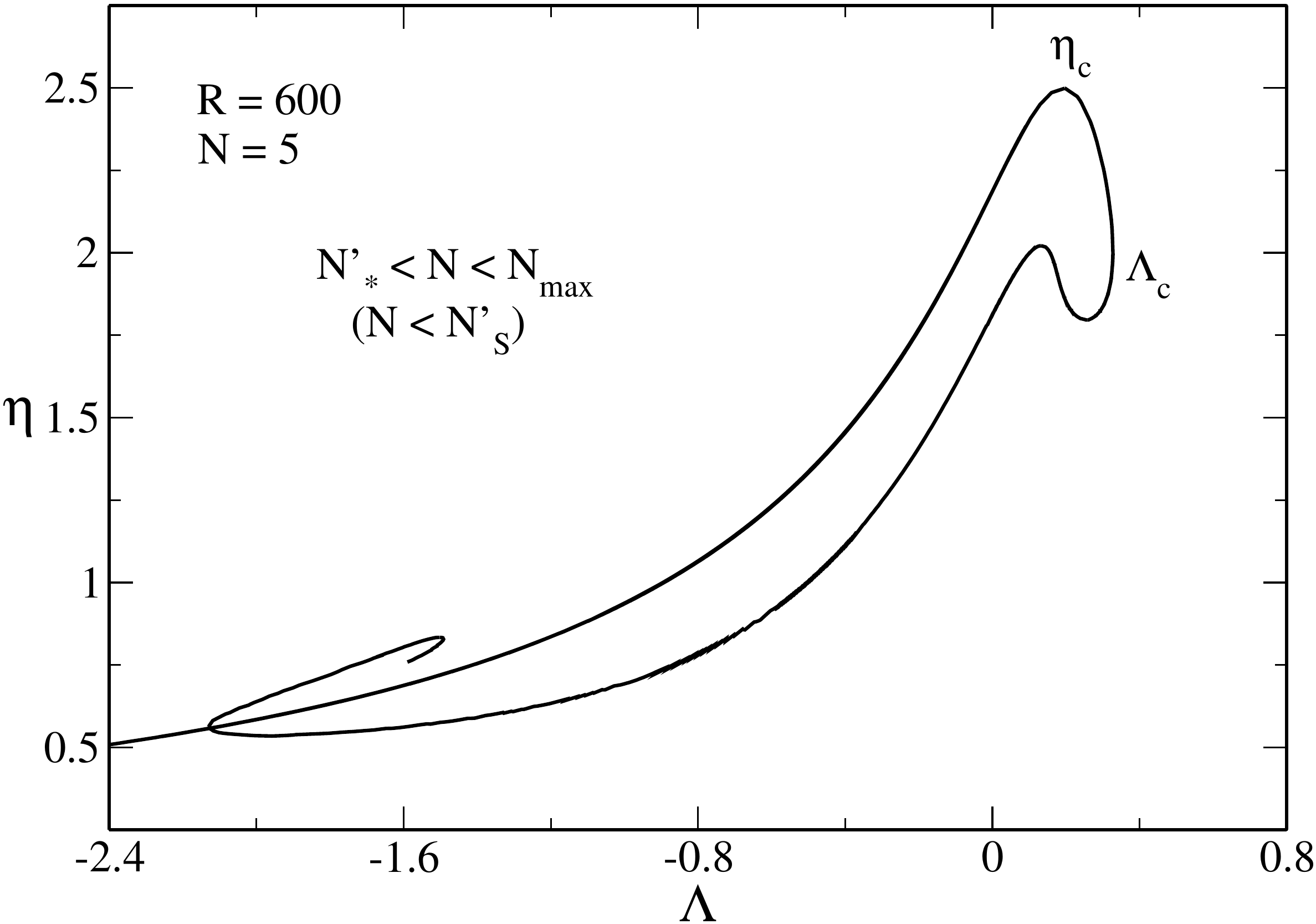}
\figsubcap{b}}
  \caption{Caloric curve of
self-gravitating fermions in general
relativity.\cite{acf} When $N>N'_*$ the condensed branch disappears completely
so that
only the collapse
at $E_c$ towards a black hole is possible.}
  \label{ac2}
\end{center}
\end{figure}

A more general study was performed 
by Alberti, Chavanis and Roupas\cite{acf,rc,calettre} who considered arbitrary
values of $R$ and $N$. When $N>N_{\rm OV}$, they identified the existence of a
new turning point of temperature in the canonical ensemble and  a
new turning point of binding energy in
the microcanonical ensemble below which the system collapses and forms a black
hole of mass $M_{\rm OV}$ (see Fig. \ref{ac1}). This is the finite temperature
generalization of the
result originally
found by Oppenheimer and Volkoff\cite{ov} at $T=0$. These results lead to the
following scenario (we restrict ourselves to the microcanonical ensemble which
is the
most relevant). At high energies, the system is in the gaseous phase. Below a
critical energy $E_c$ it becomes
thermodynamically unstable and experiences a
gravothermal catastrophe. However, core collapse stops when quantum
mechanics (Pauli's exclusion principle) comes into play. This  leads to
the formation of a fermion ball surrounded by a hot halo. Below $E''_c$, the
condensed phase  becomes
thermodynamically and dynamically unstable (in a general relativistic sense) and
collapses towards a black hole. Alberti and Chavanis\cite{acf} also evidenced a
critical
particle number $N'_*$ above which the condensed phase completely disappears
(see Fig. \ref{ac2}). In
that case,
there is no possibility to form a fermion ball. The gravothermal catastrophe at
$E_c$ directly leads to a black hole. In conclusion, for $N<N_{\rm OV}$  the
system forms a fermion ball;  for $N_{\rm OV}<N<N'_*$ the system
generically forms a fermion ball, then (possibly) a black hole; for  $N>N'_*$ 
the system directly forms a black hole.  Alberti and Chavanis\cite{acf}
emphasized the core-halo
structure of the
equilibrium states in the microcanonical ensemble and mentioned the relation to
red-giants (leading to white dwarfs) and supernovae (leading to neutron stars
and black holes) suggested in
Refs.\cite{pomeau1,supernova,supernovalettre,calettre}. They also
provided
the complete phase diagram
of the general relativistic Fermi gas.

\section{Truncated models}

The study of phase transitions in the self-gravitating Fermi gas can be 
extended to the fermionic King model described by the distribution function
\begin{equation}
\label{scl8bec}
f=A\frac{e^{-\beta(\epsilon-\epsilon_m)}-1}{1+\frac{A}{\eta_0}e^{
-\beta(\epsilon-\epsilon_m)}}\qquad (\epsilon\le\epsilon_m).
\end{equation}
The fermionic King 
model was  introduced heuristically by Ruffini and Stella\cite{stella} as a
generalization of the classical King model\cite{king}
\begin{equation}
\label{scl8becb}
f=A \left\lbrack
{e^{-\beta(\epsilon-\epsilon_m)}-1}\right\rbrack\qquad (\epsilon\le\epsilon_m).
\end{equation}
The fermionic King 
model was also introduced 
independently by Chavanis \cite{mnras} who derived it from a kinetic theory
based on the fermionic Landau equation.  The fermionic King model is
more realistic
than the usual fermionic model because it avoids the need of an artificial box
to confine
the system. The nonrelativistic fermionic King model
was studied by Chavanis {\it et al.}\cite{clm1,clm2} who showed that the density
profiles generically have a core-halo
structure with a quantum core (fermion ball) and a tidally truncated isothermal
halo leading to flat rotation curves.\footnote{The name ``fermionic King model''
was introduced in Ref. \cite{clm2}.}  They also studied the caloric curves,
the thermodynamical stability of the equilibrium states, and the phase
transitions between gaseous and condensed states (see Figs. \ref{kingc}
and \ref{kingf}). They showed that the
phenomenology of phase transitions in the fermionic King model is the same as
in the case of box-confined systems obtained in Refs.\cite{pt,ijmpb}. The
results of  Chavanis {\it et al.}\cite{clm2}  have been generalized by
Arg\"uelles {\it et al.}\cite{rarnew} to the fermionic King model in general
relativity. They
also found that the phenomenology of phase transitions in the general
relativistic fermionic King
model is the same as in the case of box-confined systems obtained in
Refs.\cite{acf,rc,calettre}.

\def\figsubcap#1{\par\noindent\centering\footnotesize(#1)}
\begin{figure}[h]%
\begin{center}
\parbox{2.1in}{\includegraphics[width=2in]{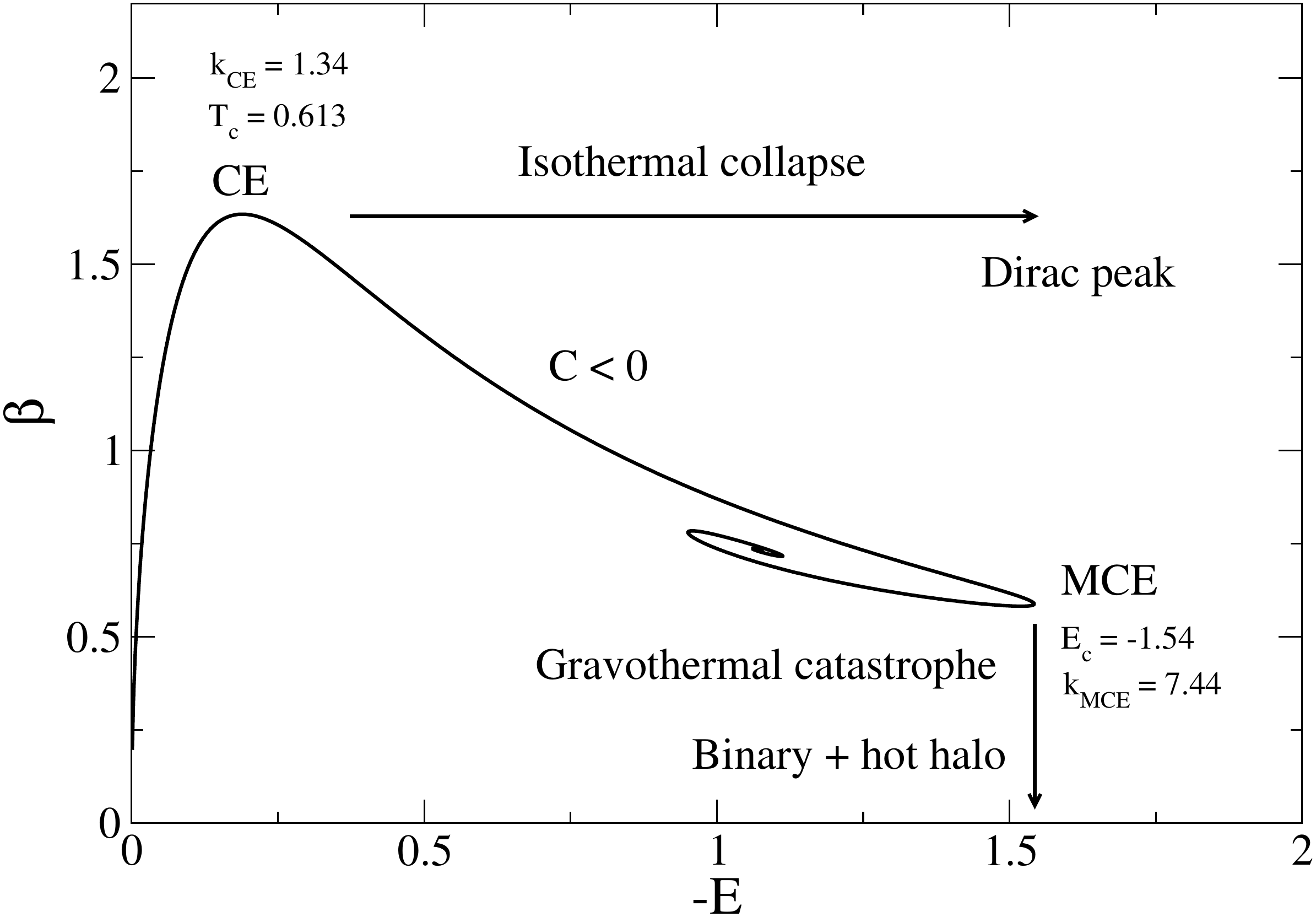}
\figsubcap{a}}
  \hspace*{4pt}
\parbox{2.1in}{\includegraphics[width=2in]{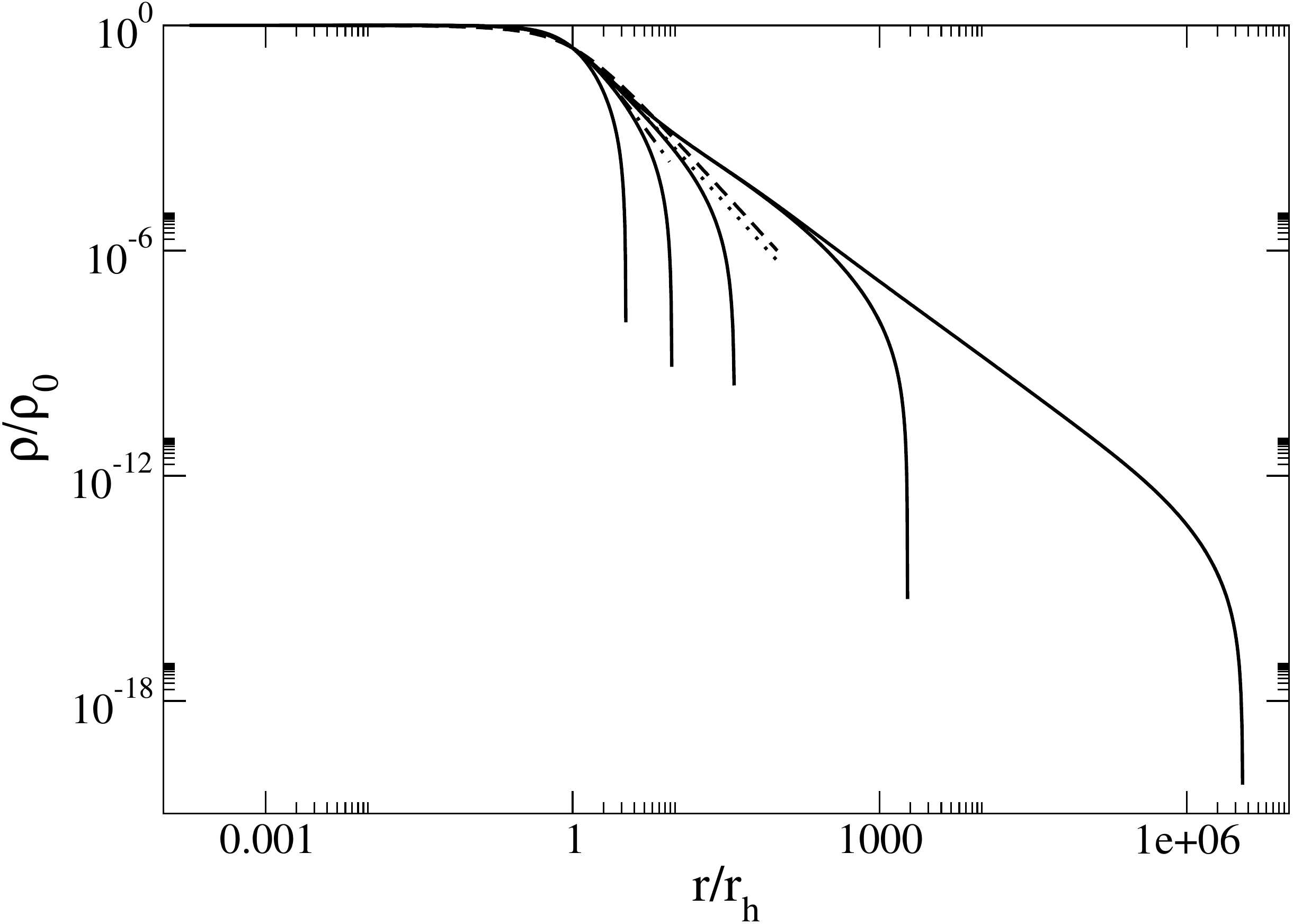}
\figsubcap{b}}
  \caption{(a) Caloric curve of the nonrelativistic classical King
model.\cite{clm1} (b) Density
profiles
along the series of equilibria.\cite{clm1} The marginal (critical) King profile
at the
point of gravothermal instability is relatively close to the
Burkert\cite{burkert} profile
(dashed line) which provides a good fit of the density profile of dark matter
halos.}
  \label{kingc}
\end{center}
\end{figure}

\def\figsubcap#1{\par\noindent\centering\footnotesize(#1)}
\begin{figure}[h]%
\begin{center}
\parbox{2.1in}{\includegraphics[width=2in]{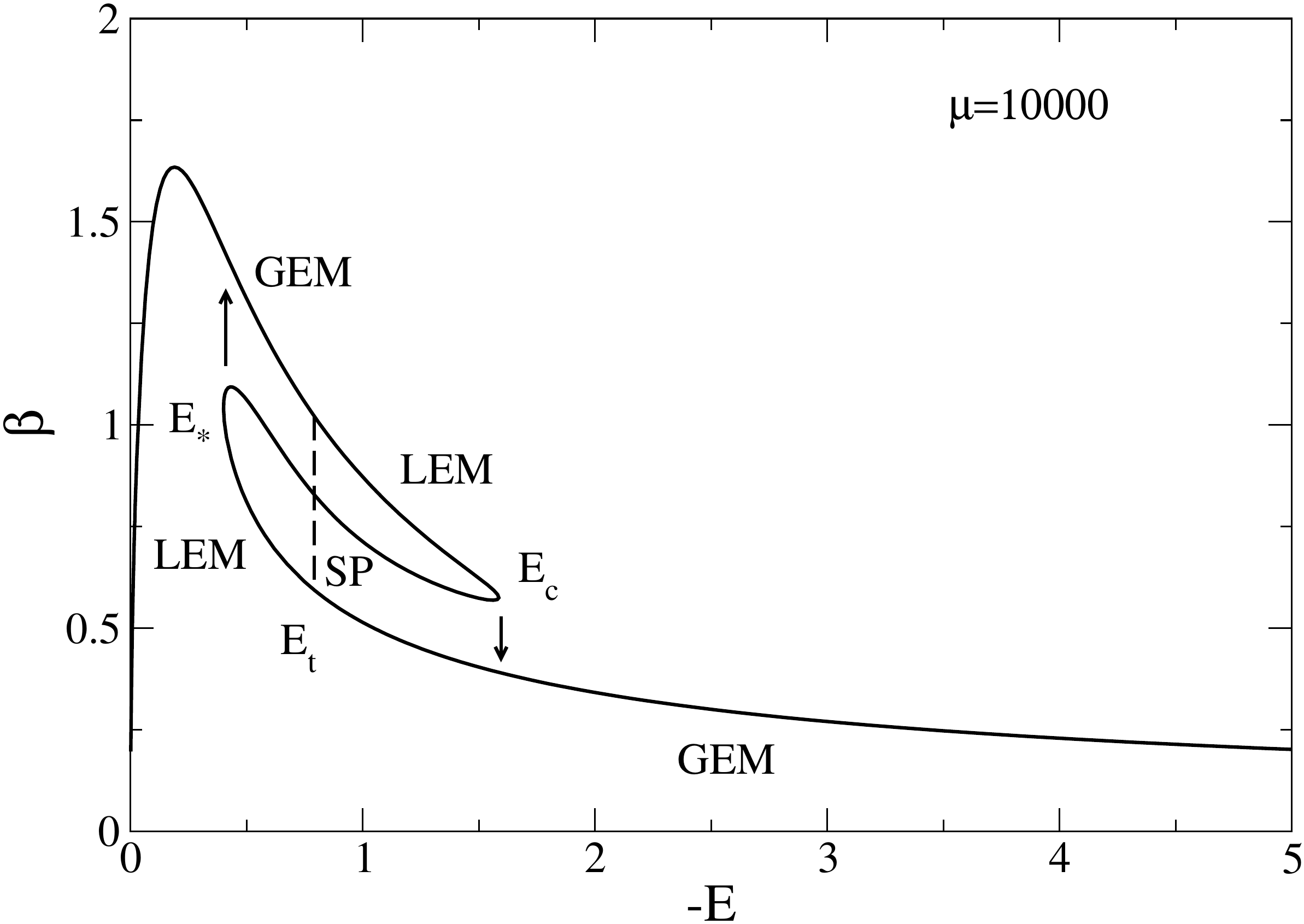}
\figsubcap{a}}
  \hspace*{4pt}
\parbox{2.1in}{\includegraphics[width=2in]{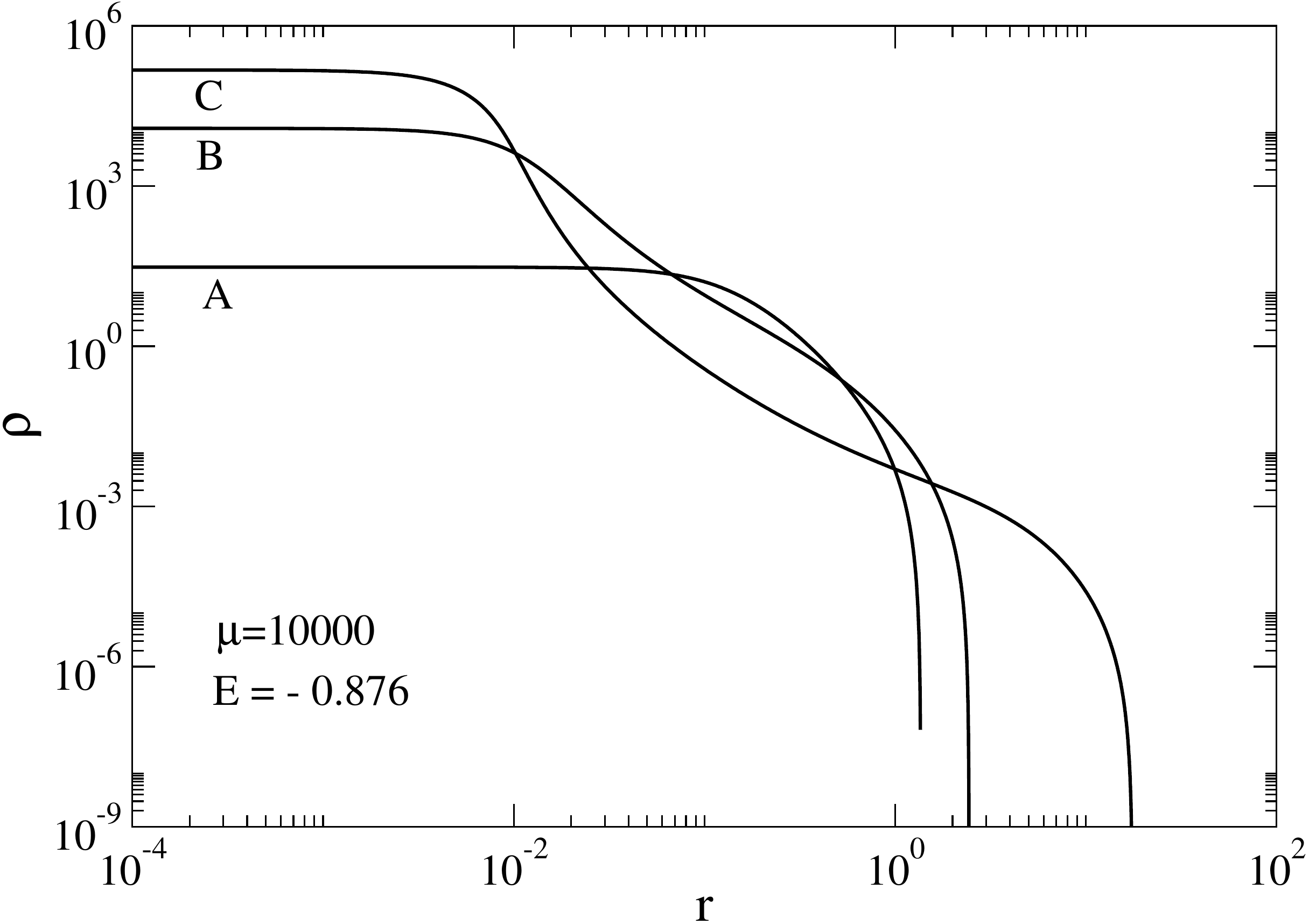}
\figsubcap{b}}
  \caption{(a) Caloric curve of the nonrelativistic fermionic
King model for large systems.\cite{clm2} (b)
Density profiles corresponding to the stable gaseous phase (A), the stable 
condensed phase (C), and the unstable intermediate  solution (B).\cite{clm2} }
  \label{kingf}
\end{center}
\end{figure}

\section{Application to dark matter halos}

In addition to white dwarfs and neutron stars,\cite{shapiroteukolsky} the
self-gravitating Fermi gas model has been applied in the context of
dark matter
halos made of massive neutrinos.  The
suggestion that dark matter is made of massive neutrinos was originally
proposed by
Markov
\cite{markov} and
Cowsik and McClelland.\cite{cmc1,cmc2} This suggestion has been
developed by numerous authors (see 
the introduction of Ref.\cite{gr1} for an exhaustive list of references). The
first models decribed dark matter halos at
$T=0$ using the  equation of state of a completely degenerate fermion gas either
in the nonrelativistic limit 
or in general
relativity.
Subsequent models considered dark matter halos at nonzero temperature showing
that they have a core-halo structure consisting of a dense core (fermion
ball) solving the core-cusp problem of classical cold dark
matter surrounded by a dilute isothermal atmosphere
leading to flat rotation curves. Most models were based on the
ordinary Fermi-Dirac distribution in Newtonian gravity or
general relativity. Some models
were based on the more realistic 
fermionic King model (describing tidally truncated fermionic dark matter 
halos).
The self-gravitating Fermi gas was also studied  in
relation to the violent relaxation of collisionless
self-gravitating systems described by the Lynden-Bell\cite{lb} distribution
which is formally similar to the Fermi-Dirac distribution. As argued in Refs.
\cite{clm1,clm2}, the theory
of violent relaxation may justify how a collisionless gas of
self-gravitating fermions, such as a dark matter halo, can reach
a statistical equilibrium state described by the Fermi-Dirac distribution on a
timescale
smaller than the age of the universe.\footnote{The relaxation time
due to close gravitational
encounters exceeds the age of the universe by many orders of magnitude. The
collisional relaxation time may be shorter if the fermions
are
self-interacting.\cite{modeldmF}}

The detailed study of the motion of S-stars near the Galactic center has
revealed the presence of a very massive central object, Sagittarius A* (Sgr A*).
This central object is usually associated with a supermassive black hole (SMBH)
of mass $M=4.2\times 10^6\, M_{\odot}$ and Schwarzschild radius $R_S=4.02\times
10^{-7}\, {\rm pc}$. Whatever the object may be, its radius must be smaller than
$R_{\rm P}=6\times 10^{-4} \, {\rm pc}$ ($R_{\rm P}=1492\, R_S$), the S2
star pericenter. Similar objects are expected to reside at the
center of most spiral and elliptical galaxies, in active galactic nuclei (AGN).
Although it is commonly believed that these objects are SMBHs, this is not yet
established on a firm
observational basis in all cases. Some authors have proposed that such objects
could be fermion balls  or
boson stars that could mimic a SMBH. Let us consider
this possibility in the framework of the fermionic dark matter model. More
precisely, let
us investigate if a fermion ball can mimic a SMBH at the
center of the Galaxy.

Bilic {\it et al.}\cite{btv} developed a general relativistic model of
fermionic dark matter halos at nonzero temperature with a fermion mass $m=15\,
{\rm
keV/c^2}$ that describes both the center and the halo of the Milky Way in a
unified manner. The density profile has a core-halo structure made of a quantum
core (fermion ball) surrounded by a classical isothermal atmosphere.  The core
and the halo are separated by an
extended plateau. By using
the usual
Fermi-Dirac distribution and choosing parameters so as to fit observational data
at large distances, they found
a fermion ball of mass $M_c=2.27\times 10^6\, M_{\odot}$ and radius
$R_c=18\, {\rm mpc}$ and argued that this fermion ball can
mimic a SMBH like Sgr A$^*$. Unfortunately, its radius is larger by a
factor $100$ than the bound
$R_P=6\times
10^{-4}\, {\rm pc}$ set by later observations. This is why
Bilic and coworkers abandoned this fermion ball scenario (R. Viollier, private
communication). The same
problem was encountered later by Ruffini  {\it et al.} \cite{rar} who developed
a similar
 model with a fermion mass $m\sim 10\, {\rm
keV/c^2}$.

More recently, Arg\"uelles {\it et al.} \cite{krut} considered the general
relativistic fermionic King model accounting for a tidal confinement.
They applied this model to the Milky Way and determined the parameters by
fitting the core-halo profile to the observations. For  a fermion mass $m=48\,
{\rm keV/c^2}$ they obtained a fermion ball of mass $M_c=4.2\times 10^6\,
M_{\odot}$ and radius
$R_c=R_P=6\times 10^{-4}\, {\rm pc}$ which, this time, is consistent with the
observational constraints of Sgr A$^*$. Therefore, in order to obtain accurate
results, it is
important to use the
fermionic King model \cite{clm2,krut} instead of the usual fermionic
model.\cite{ijmpb,btv,rar} Arg\"uelles {\it et al.} \cite{krut} managed to
fit
the entire density profile and the entire rotation curve of the Milky Way with
the 
fermionic King distribution and argued that a fermion ball can mimic the effect
of a SMBH like Sgr A$^*$. This
scenario is very attractive because it can explain the whole structure of the
galaxy, the supermassive central object and the isothermal halo, by a single
distribution (the fermionic King model\cite{stella,mnras}).

Developing the theory of phase transitions in the self-gravitating Fermi gas,
Chavanis {\it et al.} \cite{clm2} argued that the core-halo solution
considered by Bilic {\it et al.},\cite{btv} Ruffini  {\it et al.},\cite{rar}
and Arg\"uelles {\it et al.} \cite{krut} with a small fermion ball mimicking a
SMBH surrounded by a classical isothermal
atmosphere, which  was
claimed to reproduce the structure of the Milky Way, is thermodynamically
unstable because it lies in the intermediate
branch of the caloric curve between the first and the last turning points of
energy (see Fig. \ref{dp}-b). As a result, it is a saddle point of entropy, not
an entropy
maximum. Therefore, Chavanis {\it et al.} \cite{clm2} concluded that
this type of solution is not likely to result from a natural evolution and,
consequently, they questioned the possibility  that a fermion ball could  mimic
a central SMBH.

\begin{figure}[h]
\begin{center}
\includegraphics[width=2.5in]{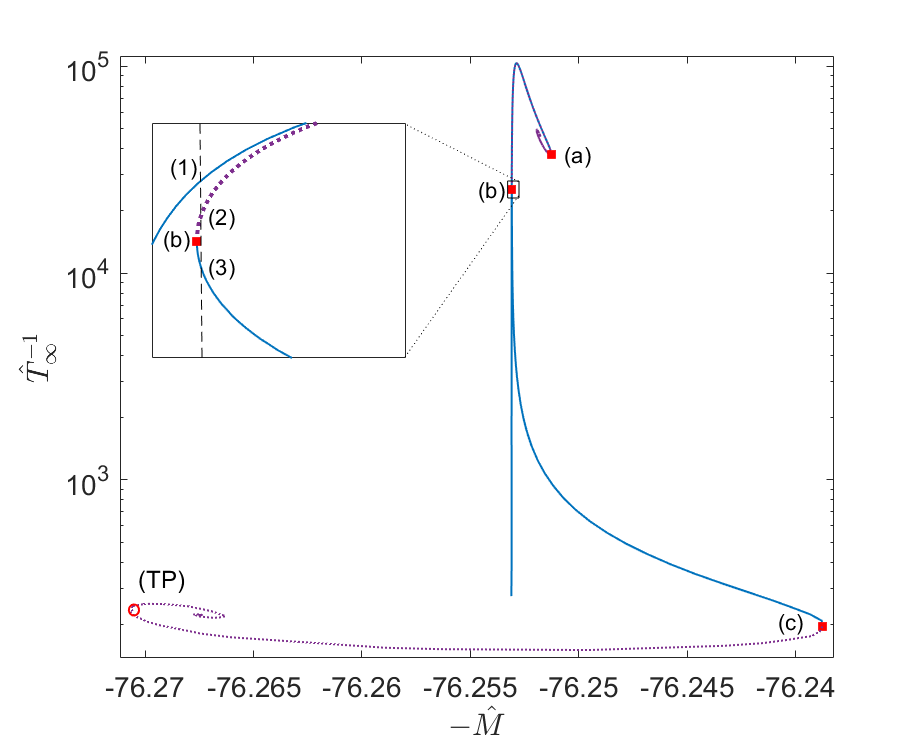}
\end{center}
\caption{Caloric curve of the general relativistic fermionic
King model (from Ref.\cite{rarnew}).}
\label{gfking}
\end{figure}

Following this study,\cite{clm2} Arg\"uelles {\it et al.} \cite{rarnew} computed
the
caloric curve 
of the fermionic  King model in
general relativity (see Fig. \ref{gfking}). They showed that the  core-halo
solution
of Ref.\cite{krut} which
provides a good agreement with the structure of the
Milky Way  lies just
{\it after} the turning point (b) of energy (see the inset of Fig.
\ref{gfking}), so that it is thermodynamically stable
in the microcanonical ensemble which is the correct ensemble to
consider.\footnote{Chavanis {\it et al.} \cite{clm2} did not focus on the
stable branch of
condensed states located after point (b) because they argued
that these solutions are not astrophysically
relevant. Indeed, by considering particular solutions of the condensed branch,
they observed that these solutions have a too extended envelope that
is not consistent with the structure of dark matter halos (see solution C in
Fig. \ref{kingf}-b and Figs. 38-45  of Ref.\cite{clm2}).
Although this claim is
correct
for most of the solutions on the condensed branch, it turns out that the
solutions located just
after point (b) {\it are} astrophysically
relevant and correspond to the core-halo solutions studied by Arg\"uelles
{\it
et al.}\cite{krut}.} This is a very
interesting result because it shows that this core-halo
structure {\it may} arise from a natural evolution in
the sense of Lynden-Bell. This gives further support to the scenario according
to which    a fermion ball could mimic a SMBH
at the center of the
galaxies. 

However, it does not prove that this core-halo structure  with a very high
central density will necessarily arise
from a natural evolution. The reason is that violent relaxation is in general
incomplete.\cite{lb} The fluctuations of the
gravitational potential that are the engine
of the collisionless relaxation can die out before the system has reached
statistical equilibrium. Therefore, it is not
clear if violent relaxation can produce
this type of structures.\footnote{It may
be
easier to form core-halo configurations with a very high central density if the
fermions are self-interacting and if the Fermi-Dirac equilibrium state results
from a collisional relaxation of nongravitational origin
instead of a collisionless relaxation as suggested in Ref.\cite{modeldmF}.} In
order to
vindicate
this scenario, the next step would be to perform direct numerical simulations
of collisionless fermionic matter to see if they spontaneously generate fermion
balls with the characteristics of SMBHs. A purely gaseous solution 
without  quantum core, which is also a
maximum entropy state, may be easier to reach through a violent relaxation
process and is consistent with the observations. However, it does not account
for a massive central object at the center of the galaxies. In that case,
we either have to introduce a primordial SMBH ``by hand'' or advocate a
scenario of gravitational collapse such as the one described
below.\cite{modeldmF}

For a fermion mass $m=48\, {\rm keV/c^2}$, the mass $M_h=10^{11}\, M_{\odot}$ of
the Milky Way is larger than the OV mass $M_{\rm OV}=2.71\times 10^8\,
M_{\odot}$, so we have to take into account general relativity  in the
caloric curve. As first shown by  Alberti and Chavanis \cite{acf,calettre} for
box-confined fermionic systems, and
recovered by Arg\"uelles {\it et al.} \cite{rarnew} for tidally
truncated models,
relativistic effects create a new turning point of energy in
the caloric
curve at which the
condensed branch terminates (see Figs. \ref{ac1}, \ref{ac2} and
point (c) in Fig.
\ref{gfking}). Below
$E''_c$ the system collapses towards
a black hole.

Suppose that violent relaxation selects the gaseous solution. 
On a secular timescale, because of collisions, the system follows the upper
branch of the series of
equilibria
(gaseous states) up to the point of minimum energy $E_c$. At that point, it
becomes
thermodynamically unstable and undergoes a gravothermal catastrophe.
However, core collapse is stopped by quantum
mechanics, leading to
the formation of a fermion ball.
Then, if the energy keeps decreasing, the system  follows the lower branch
of the series of
equilibria up to the point of minimum energy $E''_c$ where it becomes
thermodynamically and dynamically unstable (in a general relativistic sense) and
collapses towards a SMBH of mass $M_{\rm OV}$. A similar outcome arises if
violent relaxation selects
the core-halo 
solution where the fermion ball mimics a SMBH. On a secular timescale,
the system follows the lower branch of the series of equilibria up to the
point
of minimum energy $E''_c$ at which it
collapses towards a
SMBH. In the two cases, the ultimate fate of the system 
is to form a SMBH of mass $M_{\rm OV}$ surrounded by an envelope. However, the
formation of a SMBH may take time (more than the age of the universe) so that
the two objects (fermion ball or SMBH) are possible in practice.

For a fermion mass $m=48\, {\rm keV/c^2}$, the OV mass $M_{\rm OV}=2.71\times
10^8\, M_{\odot}$ is too large to account for the mass of a SMBH like Sgr A$^*$
at the center of the Milky Way. This suggests that the object at the center of
the Galaxy (Sgr A$^*$) is a fermion ball instead of a SMBH as argued by
Arg\"uelles {\it et al.}\cite{rarnew}
However, for very large halos ($N>N'_*$), it is
shown by Alberti and Chavanis \cite{acf} that the condensed branch disappears
(see panel (b) of Fig. \ref{ac2}). In that case,
there is no stable solution with a fermion ball and the system 
necessarily collapses towards a SMBH. Therefore, medium size
galaxies like the Milky Way should harbor a fermion ball of mass $M_c=4.2\times
10^6\, M_{\odot}$ while
very large galaxies like
ellipticals should harbor a SMBH of mass $M_{\rm OV}=2.71\times
10^8\, M_{\odot}$ that could even grow by accretion. This
scenario\cite{modeldmF}
could account for the mass of SMBHs in AGNs like the one
recently photographed in M87 ($M_h\sim 10^{13}\, M_{\odot}$ and $M_{\rm BH}\sim
10^{10}\, M_{\odot}$).

On the other hand, for a much larger fermion mass $m=386\, {\rm
keV/c^2}$, the OV mass $M_{\rm
OV}=4.2\times 10^6\, M_{\odot}$ is comparable to the mass of Sgr A$^*$.
Furthermore, when applied to the Milky Way, the caloric
curve corresponding to $m=386\, {\rm
keV/c^2}$ is similar
to the one reported in panel (b) of Fig. \ref{ac2} so
there is
no
possibility to form a fermion ball. In that
case, the Milky Way could have undergone a gravitational collapse at
$E_c$ leading  to
a SMBH of mass $M_{\rm OV}=4.2\times 10^6\, M_{\odot}$.\cite{modeldmF} In
this process, the
halo surrounding the SMBH is left undisturbed and could correspond to a marginal
classical King profile, which is known\cite{clm1,clm2} to give a good
agreement with the empirical Burkert\cite{burkert} profile of observed dark
matter halos
(see Fig. \ref{kingc}-b).

Different scenarios are possible depending on the
value of the fermion
mass $m$. Arg\"uelles {\it et al.} \cite{krut,rarnew}
determined the mass of the fermionic dark matter particle in such a way that the
fermion ball that would be at the center of a large fermionic dark matter halo,
like the one that surrounds the Milky Way, mimics the effect of a SMBH of mass
$M_c=4.2\times 10^{6}\, M_{\odot}$ and radius $R_c=6\times 10^{-4}\,
{\rm pc}$ like Sgr A$^*$. This leads to a fermion mass $m=48\, {\rm
keV/c^2}$.\footnote{In very recent works, Becerra-Vergara {\it
et al.}\cite{bvetal1,bvetal2} showed that the gravitational potential of a
fermion ball (with a particle mass $m=56\, {\rm keV/c^2}$) leads to a better fit
of the orbits of all the $17$ best resolved S-stars orbiting Sgr A$^*$
(including the S2 and G3 objects) with respect  to the one obtained by the
central SMBH model.}
Alternatively, Chavanis\cite{mcmh,modeldmF} determined the mass of the fermionic
dark matter
particle by
arguing that the smallest halos observed in the universe (dSphs like Fornax)
with a typical mass $M\sim 10^8\, M_{\odot}$ and a typical radius $R\sim 1\,
{\rm kpc}$ represent the ground state of the self-gravitating Fermi gas
at $T=0$. This yields a much smaller fermion mass $m=165\, {\rm eV/c^2}$. 
When this model is applied to the  Milky Way,\cite{modeldmF} it leads to a
large fermion ball of mass $M_c=9.45\times 10^9\, M_{\odot}$ and radius
$R_c=240\, {\rm pc}$. Therefore, it predicts the existence of a
large dark matter bulge at the center of the Galaxy instead of a compact fermion
ball mimicking a SMBH.\footnote{In that case, a primordial SMBH has to be
introduced ``by hand'' at the center of the Galaxy in order to account for 
the presence of Sgr
A$^*$.} A large dark matter bulge is not inconsistent with the
observations and
may even solve some issues. For example, De Martino {\it et al.} \cite{martino}
have
argued that the presence of a bosonic dark matter bulge (soliton) of mass
$M_c\simeq 10^9\, M_{\odot}$ and radius $R_c\simeq 100 \, {\rm pc}$ at the
center of the Galaxy may account
for the dispersion velocity peak observed in the Milky Way.  A large dark matter
bulge
made of fermions should have the same effect.\cite{modeldmF}

Finally, we mention potential difficulties or, alternatively, potentially
important predictions associated with the model of Arg\"uelles {\it et
al.}
\cite{krut,rarnew}. If the
fermion mass
is $m=48\, {\rm keV/c^2}$, dark matter halos of mass $M_{h}=10^8\,
M_{\odot}$ such as dSphs like Fornax should have a very pronounced core-halo
structure
since they do not correspond to the
ground state
of the self-gravitating Fermi gas (unlike the model
of Ref.\cite{modeldmF} with $m=165\, {\rm eV/c^2}$). More precisely, the
fermionic dark matter
model with
a
fermion mass $m=48\, {\rm keV/c^2}$ predicts that dSphs of mass
$M_{h}=10^8\,
M_{\odot}$ should contain a fermion ball of
mass $M_c=1.57\times 10^4\, M_{\odot}$ and 
radius $R_c=5.42\, {\rm mpc}$ possibly mimicking an intermediate mass
black hole.\cite{modeldmF} This result is
consistent with the detailed work of Arg\"uelles {\it et al.} \cite{krutmin} who
obtained dense cores of mass between $M_c=10^3\, M_{\odot}$  and  $M_c=10^6\,
M_{\odot}$ depending on the central effective temperature of the fermions.
This
is either a very important prediction (if confirmed by observations) or the
evidence that this model is incorrect (if invalidated by observations). It would
be
extremely important to clarify this issue by confronting the model of
Arg\"uelles {\it et al.} \cite{krut,rarnew} to ultracompact halos
in order to determine which of the two models (the model of Arg\"uelles
{\it et al.} \cite{krut,rarnew} with $m=48\,
{\rm keV/c^2}$ or the one developed by Chavanis\cite{modeldmF} with $m=165\,
{\rm
eV/c^2}$ or $m\sim 1\, {\rm keV}/c^2$) is the most relevant for dark matter
halos.

\section{Conclusion}

In these Proceedings, we have provided a brief history of the self-gravitating
Fermi gas in Newtonian gravity and general relativity. We have focused
exclusively on papers that discuss the caloric curves and the mass-radius
relations of the self-gravitating Fermi gas.  We have shown
how these curves become more and more complex, displaying various types of phase
transitions
and instabilities, when gravity effects, thermal effects, quantum effects,
relativity effects and tidal effects are progressively taken into account. Of
course, there
are many more
interesting papers on self-gravitating fermions that are not reviewed here. A
detailed bibliography on the subject can be found in Refs.\cite{gr1,modeldmF}
and in standard textbooks of astrophysics.

We have applied the self-gravitating Fermi gas model to dark
matter halos. The
Fermi-Dirac distribution may be justified either from the theory
of collisionless violent relaxation\cite{lb,clm2} or from a
collisional relaxation of nongravitational origin if the fermions are
self-interacting.\cite{modeldmF} If the fermions have a small mass ($m\lesssim
1\, {\rm keV}/c^2$), the caloric curve applied to the Milky Way has an $N$-shape
structure (see Fig. \ref{rfc}-b) and the equilibrium states display a large
quantum bulge of mass $M_c\sim 10^{10}\, M_{\odot}$ and radius
$R_c\sim 100\, {\rm pc}$ surrounded by an isothermal atmosphere similar to the
Burkert profile.\cite{modeldmF} If the fermions have a large mass ($m\sim 50\,
{\rm keV}/c^2$), the caloric curve has a $Z$-shape structure (see Fig.
\ref{rfm}-a). It displays a nonrelativistic turning point of energy at $E_c$
triggering the gravothermal catastrophe. For nonrelativistic fermions, the
gravothermal
catastrophe is stopped by quantum degeneracy (Pauli's exclusion
principle).\cite{csmnras} This may lead to a compact fermion ball of mass
$M_c\sim 4.2\times 10^{6}\, M_{\odot}$ and radius $R_c\sim 6\times 10^{-4}\,
{\rm pc}$ mimicking a SMBH surrounded by an isothermal atmosphere.\cite{rarnew}
When $N>N_{\rm OV}$, which is the case for the Milky Way, a new turning point
of energy appears at $E''_c$ due to general relativity (see Figs. \ref{ac1}-b
and
\ref{gfking}). It triggers a gravitational collapse towards a SMBH. This
new turning point of energy was first evidenced in
Refs.\cite{rc,calettre,acf} for box-confined fermions and confirmed in
Ref.\cite{rarnew} for the fermionic King model. The possibility to form either a
fermion ball or a SMBH at the center of the galaxies depends on the size of the
galaxy. In medium size galaxies like the Milky Way (when
$N<N'_*$) we expect to form a fermion
ball of mass $M_c\sim 4.2\times 10^{6}\, M_{\odot}$  but in large galaxies (when
$N>N'_*$) the condensed branch disappears (see Fig. \ref{ac2}-b) and the
gravothermal catastrophe necessarily results in the formation of a SMBH of
mass $M_{\rm OV}\sim 10^8\, M_{\odot}$.

It is interesting to study the effect of the dimension of space $d$ on phase
transitions in the self-gravitating Fermi gas. This is done in
Refs.\cite{wddimd,ptdimd,virialD,sgcfd}. In particular, it is shown that
fermion stars are unstable in a universe with $d\ge 4$
dimensions. In that case, quantum mechanics cannot
stabilize matter against gravitational collapse even in the 
nonrelativistic regime.\cite{wddimd,ptdimd,virialD} This is similar to a
result found by Ehrenfest \cite{ehrenfest} who considered the effect of the
dimension of space on the laws of physics and showed that planetary motion and
the Bohr atom
would not be stable in a space of dimension $d\ge 4$. Therefore, the dimension
$d=3$ of our Universe is very particular with possible implications regarding
the Anthropic Principle.

Finally, it is interesting to compare the results obtained for fermion stars
with those obtained for boson stars and self-gravitating Bose-Einstein
condensates (BECs)  (see our contribution\cite{axions} in these
Proceedings). Similarly to fermionic dark matter halos, BEC dark matter halos
also have a core-halo structure in which the ``fermion ball'' is replaced by a
``soliton''. The analogy between fermionic and bosonic dark
matter halos is discussed in Refs.\cite{mcmh,modeldmF,modeldmB}.

\begin{figure}[h]
\begin{center}
\includegraphics[width=2.5in]{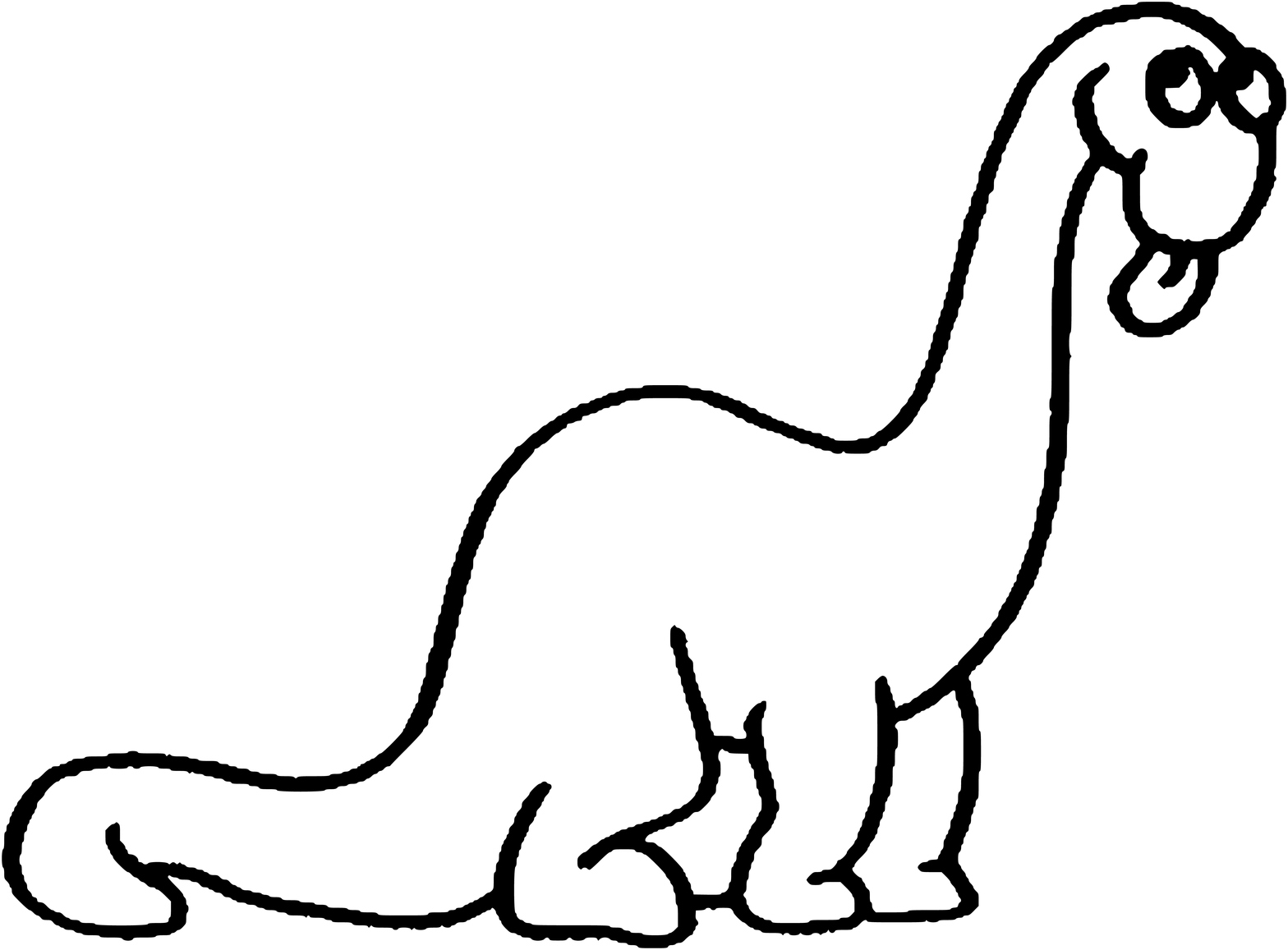}
\end{center}
\caption{Dinosaur (artistic view) similar to Fig. \ref{rfm}-a.}
\end{figure}




\end{document}